\documentclass{PoS}

\usepackage{lineno} 
\usepackage{atlasphysics}
\usepackage{subcaption}

\title{SM+Top at the LHC \\ \small\it\raggedright{Overview with a focus on Standard Model, top quark, and jet substructure results}}

\ShortTitle{SM+Top at the LHC}

\author{\speaker{T. G. McCarthy}\\
	Max-Planck-Institut f$\ddot{\textrm{u}}$r Physik, F$\ddot{\textrm{o}}$hringer Ring 6, 80805 M$\ddot{\textrm{u}}$nchen, Germany\\
        On behalf of the ATLAS, CMS and LHCb Collaborations\\
        E-mail: \email{tmccarth@mpp.mpg.de}}

\abstract{These proceedings highlight a selection of recent results by the ATLAS, CMS and LHCb collaborations.  The majority of the featured analyses make use of the large set of $\rts=13$ \TeV \ proton-proton collision data collected during the successful second run of the LHC.  A particular focus is placed on analyses of Standard Model processes involving either hadronic jets or $\Wboson$ / $\Zboson$ bosons. Searches and cross-section measurements involving top quark signatures are also given prominence, as are those targeting highly boosted objects such as $\Wboson/\Zboson$ and $\Hboson$ bosons, and which feature the use of large-radius jets and substructure techniques.}

\FullConference{An Alpine LHC Physics Summit (ALPS2018)\\
		15-20 April, 2018\\
		Obergurgl, Austria}

\begin{document}
\noindent\rule[0.5ex]{\linewidth}{1pt}
\section{Overview}

The successful operation of the LHC and its flagship experiments in recent years has provided physicists with a wealth of data from the substantial numbers of recorded proton-proton ($p$-$p$) collisions.  This has opened up the possibility to explore previously inaccessible regions of phase space in relatively common and well understood Standard Model (SM) processes, and to search for extremely rare SM processes or hints of beyond the Standard Model (BSM) processes.  This document corresponds to an overview talk which highlighted a number of selected recent searches and measurements by the ATLAS, CMS and LHCb collaborations \cite{ATLASExperiment, CMSExperiment, LHCbExperiment}.  The presented analyses are grouped into three categories, which recognizably overlap to some degree: measurement of SM processes in extreme regions of phase space and searches for rare SM processes, particularly those featuring jets and \Wboson \ or \Zboson \ bosons; measurements based on top quark signatures and searches for rare top quark processes; and searches featuring boosted SM objects which benefit from a detailed knowledge of jet substructure.

\noindent\rule[0.5ex]{\linewidth}{0.5pt}
\section{Recent Standard Model Results}



\subsection{Jet and Dijet Cross Sections}

Jet and dijet cross-section measurements allow for a good test of perturbative QCD (pQCD) in $p$-$p$ collisions.  The ATLAS collaboration performed such doubly differential measurements using a total integrated luminosity of up to 3.2 \ifb \ at a centre-of-mass energy of $\rts=13$ \TeV \ \cite{ATLAS1}.  Reconstructed jets were formed using the $\akt$ algorithm with a distance parameter $R=0.4$ \cite{antiktalgo}, and using EM-scale calorimeter clusters as the inputs.  The inclusive-jet results cover a wide kinematic range in jet transverse momentum (\pT) from 100 \GeV \ to 3.5 \TeV \ and for several separate ranges of jet rapidity up to $|y|=3.0$. In the case of the dijet cross-section measurements, the results are presented as a function of dijet invariant mass ($\mass{\rm j}$) in the range $300 \ \GeV < \mass{\rm j} < 9 \ \TeV$, and for values of the quantity $y^{*} = \frac{1}{2}|y_1-y_2|$ -- half the absolute rapidity difference between the two leading jets\footnote{The Lorentz-invariant quantity $y^{*}$ corresponds to the absolute rapidity of each of the two jets in the dijet rest frame.} -- up to $y^{*}=3.0$.  The choice of bin sizes (for \pT, $y$, $\mass{\rm j}$ and $y^{*}$) was motivated by the respective detector resolution.  The measured results from data were compared to next- and in some cases next-to-next-to-leading order (NLO and NNLO, respectively) pQCD predictions, and in all cases were found to be in good agreement with the SM predictions.  The inclusive-jet and dijet results are summarized in Figure~\ref{Fig-ATLAS-1}, where the usual convention is adopted of applying multiplicative offset factors to the results in various $|y|$ ($y^{*}$) bins in order to allow for a better visual comparison of the shapes.  Doubly and triply differential jet cross-section measurements were also performed by the CMS collaboration \cite{CMS21,CMS22,CMS23}.

\begin{figure}[!htbp]
\begin{subfigure}{0.47\textwidth}
\begin{centering}
\includegraphics[scale=0.33]{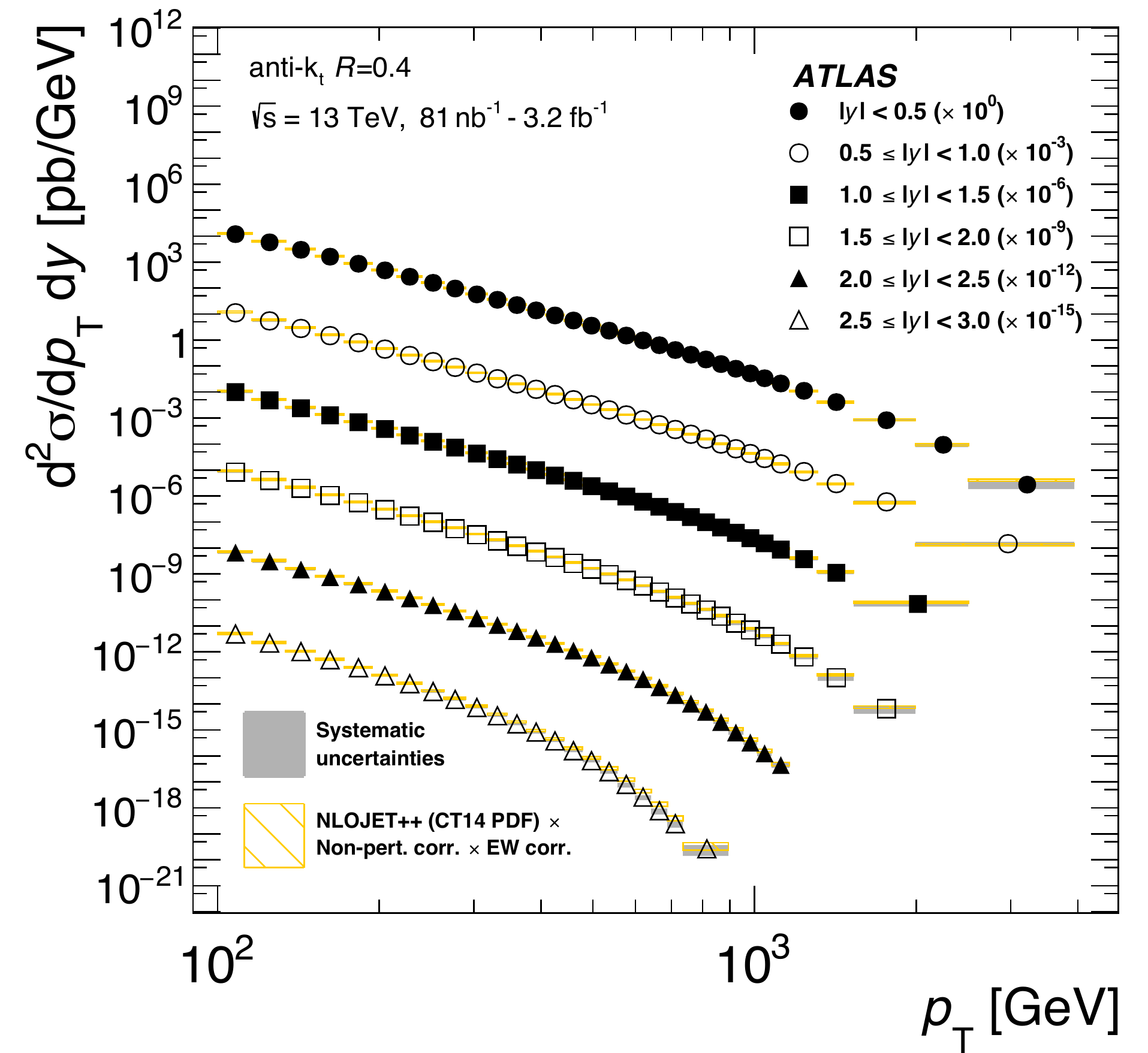}
\caption{}
\label{Fig-ATLAS-1-a}
\end{centering}
\end{subfigure} 
\begin{subfigure}{0.47\textwidth}
\begin{centering}
\includegraphics[scale=0.33]{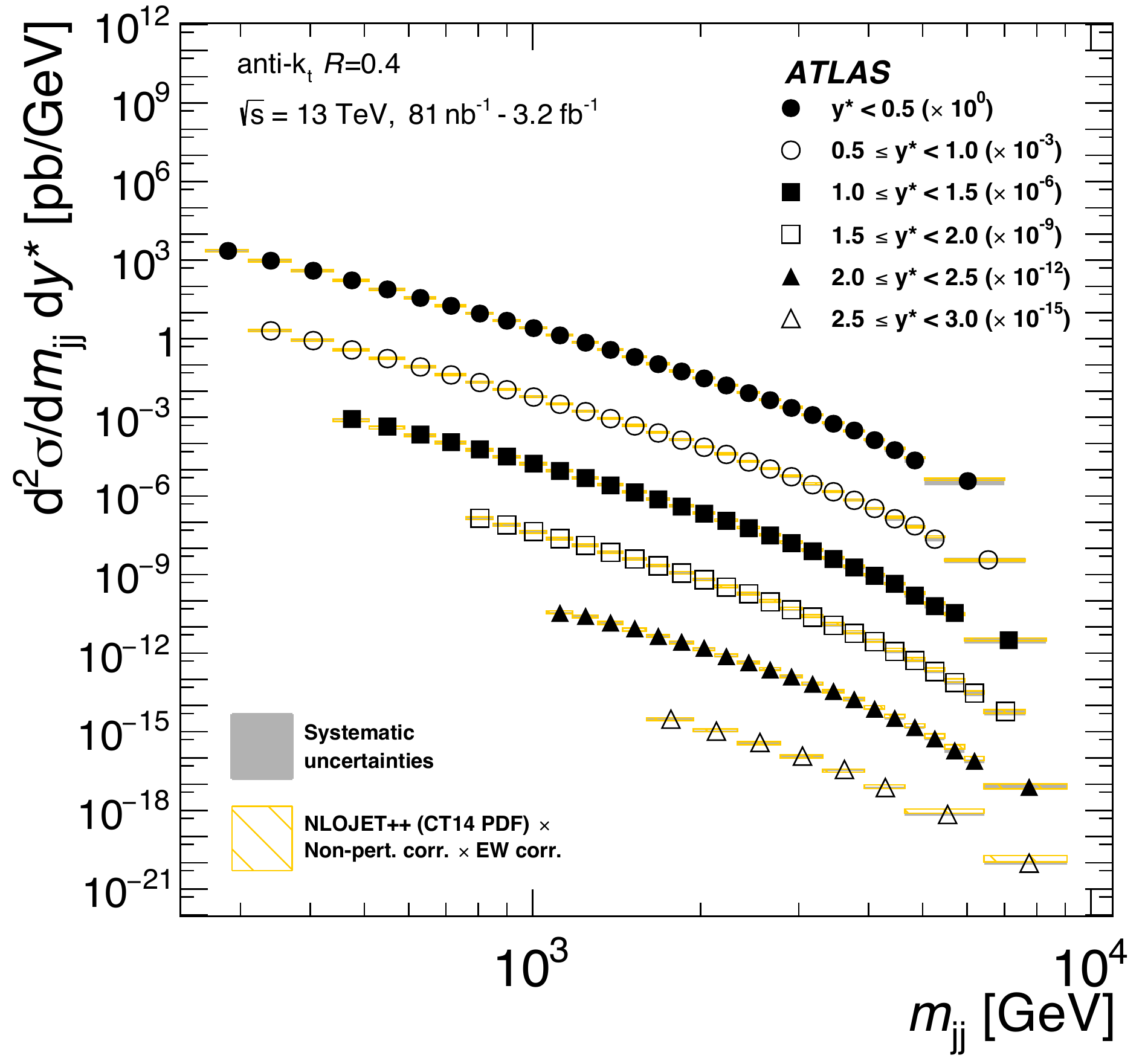}
\caption{}
\label{Fig-ATLAS-1-b}
\end{centering}
\end{subfigure} 
\caption{Results of the measured (a) jet and (b) dijet differential cross-section results at $\rts = 13$ \TeV \ by the ATLAS collaboration \cite{ATLAS1}.}
\label{Fig-ATLAS-1}
\end{figure}

\subsection{Results Featuring $\Wboson$ Bosons}

Recent analyses have been performed by the ATLAS and CMS collaborations to test SM predictions for processes involving heavy, charged electroweak (EW) bosons, and notably for rarer SM processes involving $\Wboson\Wboson$ pairs which have only recently become accessible for $p$-$p$ collisions.


The CMS collaboration has performed a set of differential $\Wboson$+jets cross-section measurements with 2.2 \ifb \ of Run 2 data -- the first such measurements to be performed with a $\rts = 13$ \TeV \ dataset \cite{CMS28}.  The analysis strategy targeted events with a single, isolated, high-\pT \ muon and considered jet multiplicities up to $N_{\rm jets} = 6$.  Differential results unfolded to stable particle level were presented as a function of a number of kinematic observables including: the transverse momentum and absolute rapidity of the $i^{\rm th}$ leading jet; the azimuthal separation between the $i^{\rm th}$ leading jet and the selected reconstructed muon\footnote{In all above-mentioned cases, the values $i = 1,2,3,4$ are considered.}; the jet multiplicity; the scalar \pT \ sum of reconstructed jets (\HT); and the separation in $\eta$-$\phi$ space between the reconstructed muon and its closest reconstructed jet.  The final results as a function of $\HT$ for the $N_{\rm jets} \geq 1$ case can be seen in Figure~\ref{Fig-CMS-28}. The measurements from data were compared to the predictions from a variety of simulated datasets.  In all cases the results for the angular observables were found to be well described.  Disagreements were observed between the data and a leading-order (LO) MG\_aMC@NLO prediction, which underestimates the measured data at low \HT, as well as at low to moderate jet \pT.


A more rare SM process features the EW production of a same-sign \Wboson \ boson pair ($\Wboson^{\pm}\Wboson^{\pm}$).  An observation of this process was recently made by the CMS collaboration in an analysis targeting events with two same-sign leptons ($ee/e\mu/\mu\mu$) and two hadronic jets \cite{CMS24}.  The signal-to-background ratio ($S/B$) was enhanced by means of cuts applied to several discriminating kinematic variables, including a requirement on the missing transverse energy (\met > 40 \GeV), where the latter is motivated by the presence of two final-state neutrinos from the leptonic \Wboson \ boson decays.  In the $e^{\pm}e^{\pm}$ channel, the probability to misidentify the charge of one of the reconstructed leptons is non-negligible, which could lead to an enhancement of any SM backgrounds featuring leptonically decaying $\Zboson$ bosons; a  \Zboson-mass window criterion (requiring selected events to satisfy $|\mass{\ell}-\mZ| > 15 \ \GeV$) was consequently used to further suppress such SM background contributions in the final set of candidate events.  The final distribution of the invariant mass of the dijet system (\mass{\rm j}) for this analysis is shown in Figure~\ref{Fig-CMS-24}, where one can observe both a good level of agreement between data and the stacked SM predictions, as well as a clear presence of the rare EW $\Wboson^{\pm}\Wboson^{\pm}$ process in the measured data.  The observation of this rare process offers a stringent test of the SM and is important in its own right; moreover the results prove useful in searches for doubly charged Higgs bosons ($\Hboson^{\pm\pm}$), for which the SM EW $\Wboson^{\pm}\Wboson^{\pm}$ production is one of the primary backgrounds. Limits at the 95\% C.L. were placed on both the cross section times branching ratio $\sigma_{\rm VBF}(\Hboson^{\pm\pm}) \times B(\Hboson^{\pm\pm}\to\Wboson^{\pm}\Wboson^{\pm})$ and on the coefficients from a set of dimension-8 operators in a modified Lagrangian in the context of an effective field theory (EFT) approach.

\begin{figure}[!htbp]
\begin{subfigure}{0.47\textwidth}
\begin{centering}
\includegraphics[scale=0.35]{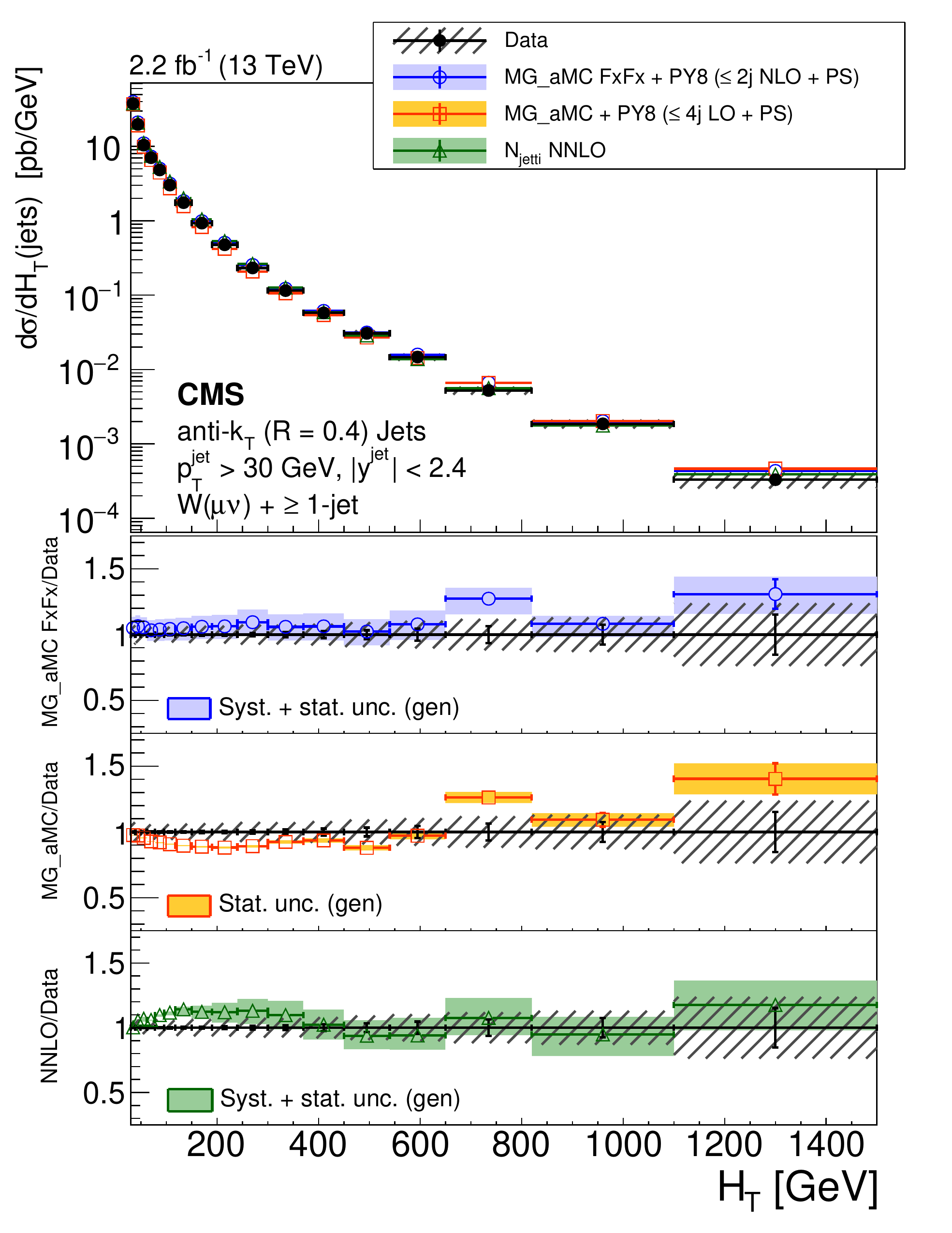}
\caption{}
\label{Fig-CMS-28}
\end{centering}
\end{subfigure} 
\begin{subfigure}{0.47\textwidth}
\begin{centering}
\vspace{2.3cm}
\includegraphics[scale=0.35]{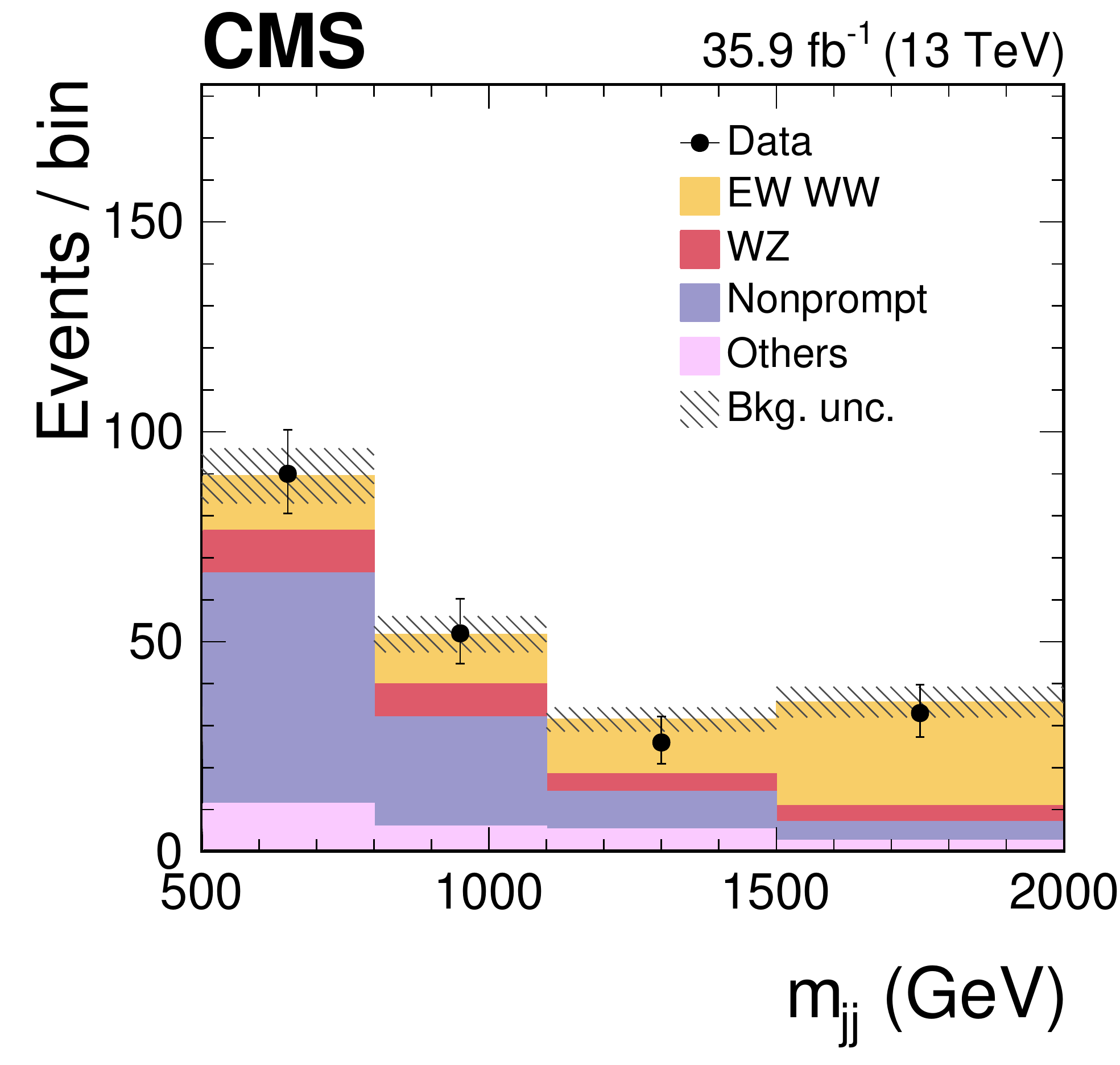}
\caption{}
\label{Fig-CMS-24}
\end{centering}
\end{subfigure} 
\label{Fig-CMS-28-CMS-24}
\caption{(a) One of the differential cross-section results from the CMS $\Wboson+$jets analysis, shown here specifically for the $N_{\rm jets} \geq$1 case as a function of the scalar \pT \ sum (\HT) \cite{CMS28} and (b) the final distribution of the invariant dijet mass system in the same-sign $\Wboson^{\pm}\Wboson^{\pm}$ analysis by CMS \cite{CMS24}.}
\end{figure}


The analogous opposite-sign process ($pp\to\Wplus\Wminus$) is far more common by comparison and represents a large source of background for many Run 2 measurements at the LHC.   An inclusive cross-section measurement of this process was performed by the ATLAS collaboration in the $e\mu$ channel with 3.2 \ifb \ of $\rts = 13$ \TeV \ data \cite{ATLAS5}.  Several resonant and non-resonant production modes were considered. The result, following an extrapolation from the fiducial to full phase space based on simulation, is a measured cross section in agreement with the value based on theoretical predictions\footnote{References for the various theoretical predictions (of order $\alphas^2$, $\alphas^3$ and $\alphas^5$) are provided in the ATLAS publication.} encompassing processes at varying orders of \alphas.  A further comparison between the measured fiducial cross sections at different centre-of-mass energies was performed, yielding: $\frac{\sigma_{pp\to\Wplus\Wminus\to e\nu\mu\nu}(\rts=13 \ \TeV)}{\sigma_{pp\to\Wplus\Wminus\to e\nu\mu\nu}(\rts=8 \ \TeV)} = 1.41 \pm \ 0.06 \ ({\rm stat.}) \pm \ 0.16 \ ({\rm syst.}) \pm \ 0.04 \ ({\rm lumi.})$.  The precision of the measured result was limited by systematic uncertainties, specifically those associated with jet reconstruction efficiencies, the jet energy scale, and the jet energy resolution.



\begin{figure}[!htbp]
\begin{subfigure}{0.47\textwidth}
\begin{centering}
\includegraphics[scale=0.67]{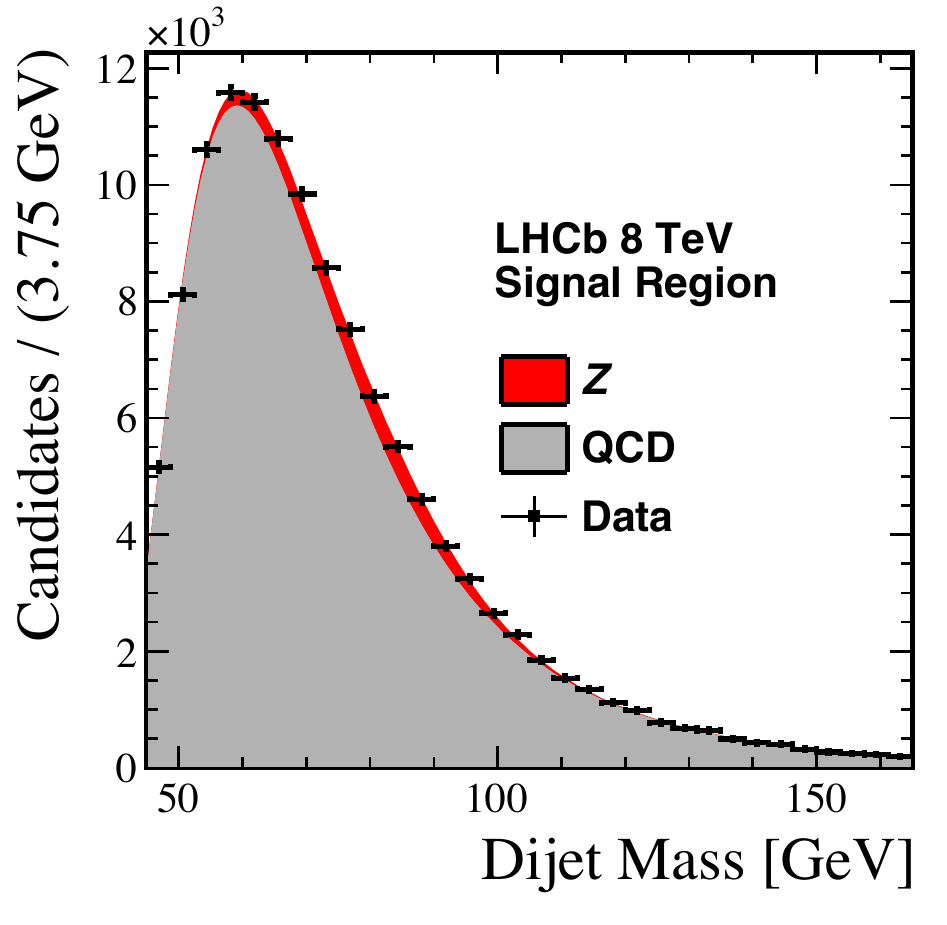}
\caption{}
\label{Fig-LHCb-43-a}
\end{centering}
\end{subfigure} 
\begin{subfigure}{0.47\textwidth}
\begin{centering}
\includegraphics[scale=0.59]{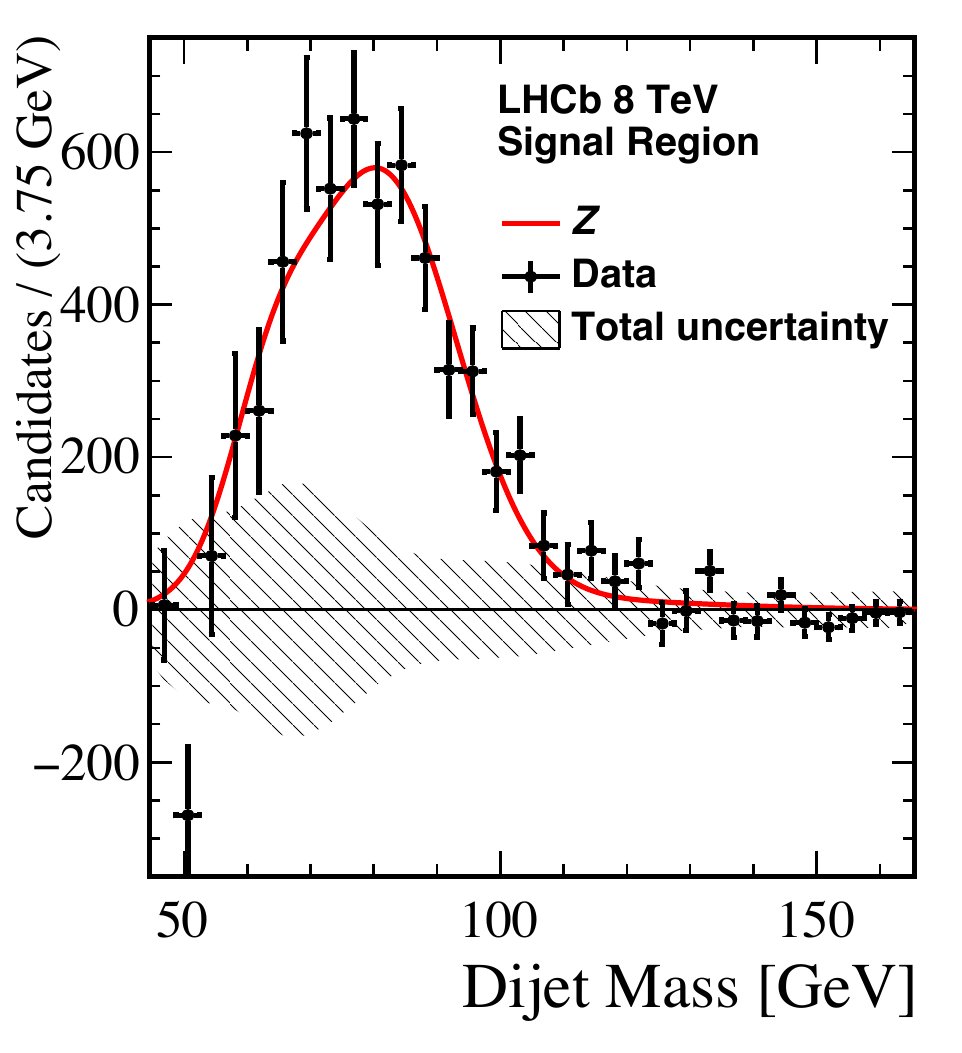}
\caption{}
\label{Fig-LHCb-43-b}
\end{centering}
\end{subfigure} 
\caption{The invariant dijet mass distribution in the LHCb search for forward $\Zboson\to\bbbar$ in (a) the final signal region and (b) the final signal region after a subtraction of the background \cite{LHCb43}.}
\label{Fig-LHCb-43}
\end{figure}

\subsection{Results Featuring $\Zboson$ Bosons}

A number of analyses have also been carried out at the LHC to study heavy, neutral EW gauge bosons.  One such analysis performed by the LHCb collaboration involved the first observation of forward-produced $\Zboson$ bosons decaying to a $\bbbar$ final state \cite{LHCb43}.  The forward tracking ability of the LHCb detector offers a unique opportunity to study such events in a non-standard kinematic regime.  The measurement was performed with Run 1 $\rts=8$ \TeV \ data. Motivated by the LO SM $t$-channel production mechanism, which features a $\Zboson$ boson recoiling against a quark or anti-quark, selected events were required to include a pair of heavy-flavour-tagged jets with an invariant mass in the range $45 < \mass{\rm j} < 165$ \GeV, and an additional reconstructed jet satisfying $\pT>10$ \GeV \ and $2.2 < \eta < 4.2$.  A Boosted Decision Tree (BDT) was trained in order to offer better discrimination between the signal and the largely irreducible QCD multijet background.  The final signal significance was at the 6$\sigma$ level, allowing for a cross-section times branching ratio measurement within the fiducial region.  A clear presence of the $\Zboson$ signal is visible in the final distributions shown in Figure~\ref{Fig-LHCb-43}, particularly after the substraction of the QCD background.  The limited statistics and uncertainties in the heavy-flavour-tagging efficiency were the dominant sources of uncertainty.  An improved understanding of $\Zboson\to\bbbar$ processes will lead to improvements in searches which feature Higgs bosons decaying similarly via $\Hboson\to\bbbar$. Moreover, in the future such measurements by LHCb with increased statistics would offer the possibility to investigate the $\Zboson$ central-forward asymmetry as a precision test of EW predictions of the SM, offering complementarity to the ATLAS and CMS measurements for which no such forward tracking is currently available. 


The CMS collaboration recently published the results of a search for $p$-$p$ collision events at $\rts = 13$ \TeV \ in which a \Zboson \ boson was produced in association with two jets via EW interactions \cite{CMS25}.  The analysis targeted leptonically decaying $\Zboson$ bosons ($\Zboson\to\ell^{+}\ell^{-}$, $\ell=e,\mu$).  A likelihood discriminant was trained to discriminate between quark- and gluon-initiated jets, the latter representing a substantial source of background.  Additional discriminating variables were used to train a BDT in order to enhance the signal significance.  The output BDT distribution for one of the final signal regions is shown in Figure~\ref{Fig-CMS-25}; in the rightmost bins with the largest $S$/$B$ one can note the clear presence of the signal.  In addition to the cross-section measurement within a fiducial kinematic region\footnote{The cross section was performed in the fiducial kinematic region in which the invariant mass of the lepton pair satisfied $\mass{\ell}>50 \ \GeV$, and the jets satisfied both a transverse momentum requirement $\pT^{j} > 25 \ GeV$ and an invariant mass requirement $\mass{\rm j} > 120 \ \GeV$.}, detailed studies were performed to investigate jet activity in the final signal region.  In all cases good agreement between data and prediction was observed.

\begin{figure}[!htbp]
\begin{subfigure}{0.47\textwidth}
\begin{centering}
\includegraphics[scale=0.35]{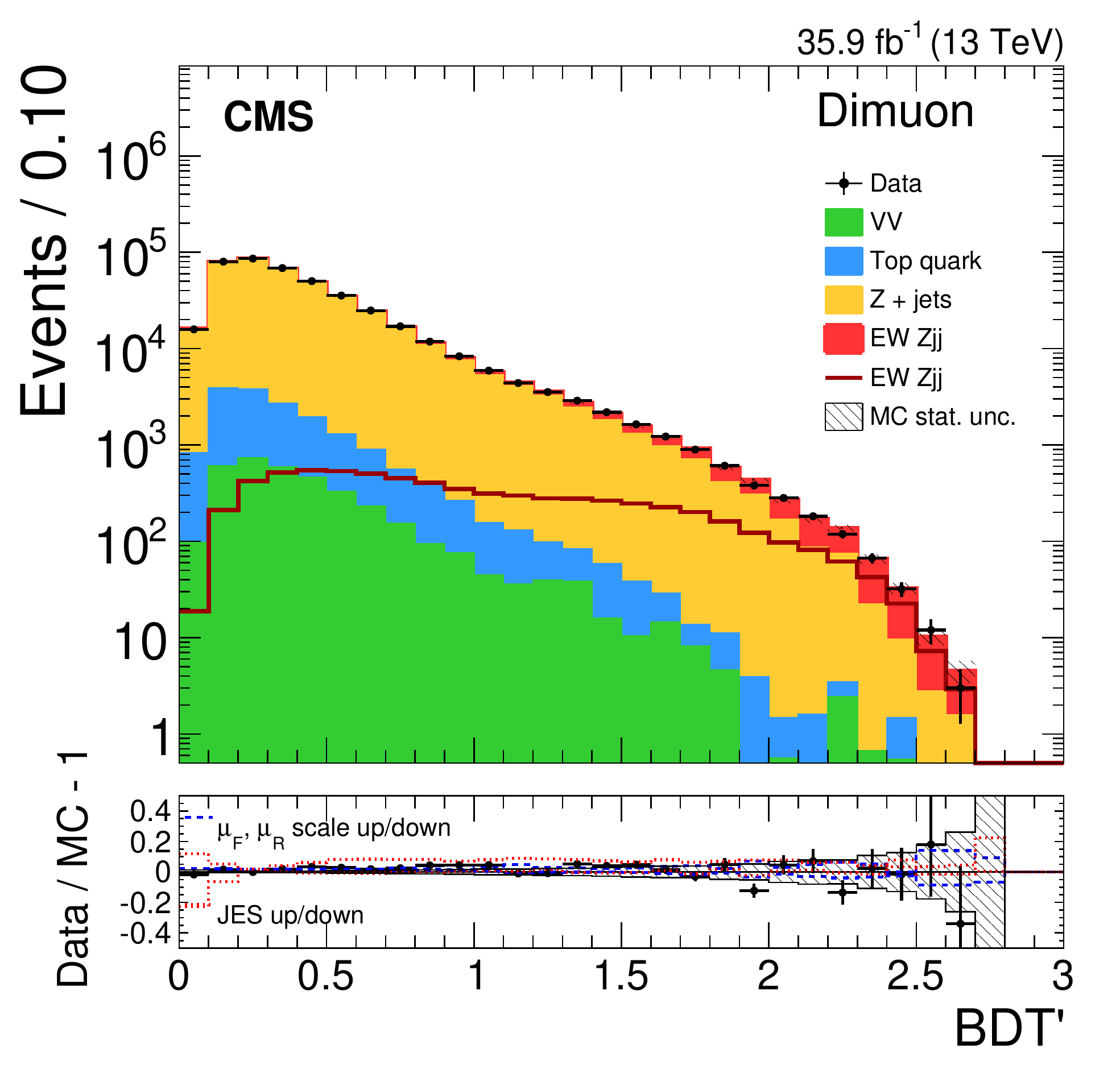}
\caption{}
\label{Fig-CMS-25}
\end{centering}
\end{subfigure} 
\begin{subfigure}{0.47\textwidth}
\begin{centering}
\includegraphics[scale=0.35]{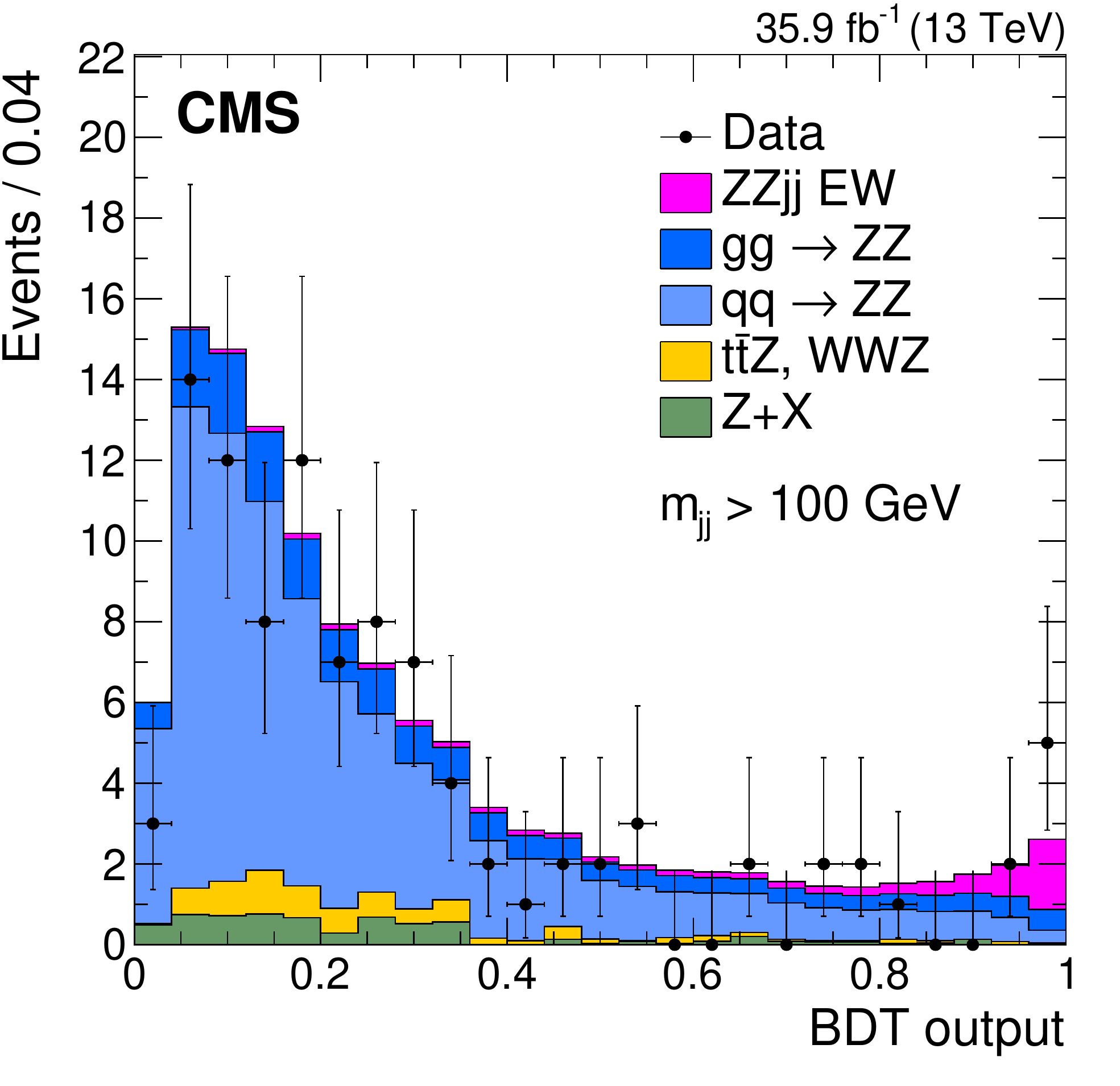}
\caption{}
\label{Fig-CMS-27}
\end{centering}
\end{subfigure} 
\label{Fig-CMS-25-CMS-27}
\caption{Final output BDT distributions for the searches by CMS for (a) the associated EW production of a $\Zboson$ boson and two additional jets \cite{CMS25} and (b) the EW production via VBS of a $\Zboson\Zboson$ pair and two additional jets \cite{CMS27}.}
\end{figure}


CMS also recently published a measurement of both the inclusive and differential production cross sections for the process $pp\to\Zboson\Zboson$ based on $\rts=13$ \TeV \ data \cite{CMS26}.  The signal processes considered include both on- and off-shell \Zboson \ bosons\footnote{The general signal process considered was $pp\to\left(\Zboson/\gamma^{*}\right)\left(\Zboson/\gamma^{*}\right)\to4\ell$.}, and the analysis strategy focused on 4$\ell$ final states ($\ell=e,\mu)$, which offer a clean selection with relatively low backgrounds.  The $gg$ and the more dominant $\qqbar$ production modes are displayed separately in final distributions.  Both the inclusive and the differential cross-section mesurements were found to be in good agreement with NLO SM predictions. Additional constraints were able to be placed on anomalous, non-SM $\Zboson\Zboson\Zboson$ and $\Zboson\Zboson\gamma$ couplings, and a measurement of the branching ratio of the rare decay mode $\Zboson\to\ell^{+}\ell^{-}\gamma^{*}\to\ell^{+}\ell^{-}\ell^{\prime+}\ell^{\prime-}$ was performed, yielding a value in agreement with the SM.  Events featuring this rare decay mode are useful in the context of mass calibration for $\Hboson\to\Zboson\Zboson^{*}\to4\ell$ decays.



A pair of $\Zboson$ bosons can also be produced in association with two hadronic jets according to the SM via so-called Vector Boson Scattering (VBS) through EW interactions -- a process featuring a quartic $\Wboson\Wboson\Zboson\Zboson$ coupling in one of the leading-order diagrams.  The CMS collaboration recently published the results of the first search for such a process \cite{CMS27}.  Selected events were required to have two $R=0.4$ particle-flow jets \cite{CMSPFlowJets} and two pairs of reconstructed leptons originating from the $\Zboson\Zboson\to\ell^{+}\ell^{-}\ell^{\prime+}\ell^{\prime-}$ decay (where $\ell, \ell^{\prime} = e,\mu$).  The $\Zboson\Zboson$ signal from the previously mentioned analysis here takes on the role of the dominant background.  Several constraints on the invariant mass of reconstructed $\Zboson$ bosons ($\mass{\ell}$), proved useful in suppressing some of the non-$\Zboson$ backgrounds, including a requirement that the pairs satisfy $\mass{\ell}>4$ \GeV \ in order to suppress the contribution from hadron decays.  The signal process is highly limited statistically; a BDT employing seven discriminating variables was used to improve the ability to discriminate between signal and background events.  Only a handful of events remain after the various requirements, but the presence of the signal is apparent in the final distribution, shown in Figure~\ref{Fig-CMS-27}, of the BDT output.  The resulting measured value of the cross section was found to be in agreement with the SM prediction.  In addition constraints were able to be placed on non-SM, so-called anomalous quartic couplings.


Several analyses featuring one or two $\Zboson$ bosons were performed by the ATLAS collaboration in addition to those mentioned above\footnote{See for example \cite{ATLAS2,ATLAS3,ATLAS4}.}, but were the focus of another dedicated contribution to this conference.

\noindent\rule[0.5ex]{\linewidth}{0.5pt}
\section{Recent Top Quark Results}

Analyses involving top quarks -- the heaviest of all fundamental particles in the SM -- allow for rigorous tests of recent NNLO and NNLL pQCD predictions.  An unprecedented number of top quarks have been produced at the centres of the main LHC experiments in recent years; previously statistically limited searches for rare SM processes featuring top quarks are now within reach based on the $\rts=13$ \TeV \ Run 2 data collected so far.   For some less rare and previously measured processes, such as the differential \ttbar \ cross-section, statistical limitations are now being relegated to the extremes of kinematic regions of phase space, where deviations from the SM may eventually be exhibited.  One set of rare processes of particular interest involves the associated production of \ttbar \ pairs with $\Wboson$/$\Zboson$ or $\Hboson$ bosons.
Moreover, with the top quark as the heaviest of all SM particles, techniques to identify top quarks with highly collimated decay products could prove vital in searches for new physics: any enhancement in the rates of boosted top quarks could provide hints of as-of-yet undiscovered BSM particles.  Finally, SM processes featuring top quarks often represent the dominant source of background to many searches for BSM physics; a better understanding of the kinematics of top quark events in general will lead to improved sensitivity in such searches.


\subsection{Standard Differential $\ttbar$ Cross-Section Results}

Aside from providing information about process kinematics and offering sensitivity to fundamental parameters of QCD, \ttbar \ cross-section measurements allow for improvements in the background modelling for LHC searches, specifically those involving signal processes for which top-antitop quark pairs represent a significant source of background.

Despite its lower relative branching ratio, the dileptonic \ttbar \ decay channel offers the advantage of very low backgrounds.  As the LHC datasets have grown, the \ttbar \ analyses have become far less statistically limited; this has allowed \ttbar \ cross-section measurements in the dileptonic channel to produce by far the more precise results.

A differential \ttbar \ cross-section measurement was recently performed by the ATLAS experiment in the dileptonic $e^{\pm}\mu^{\mp}$ channel as a function of a number of kinematic observables: the transverse momentum and absolute rapidity of the top quark ($\pT^{t}$, and $|y^{t}|$); and the transverse momentum, the absolute rapidity, and the invariant mass of the reconstructed \ttbar \ system ($\pT^{\ttbar}$, $|y^{\ttbar}|$, and $\mass{}^{\ttbar}$, respectively) \cite{ATLAS6}.  The final distributions were unfolded, based on a fiducial volume, from detector to particle level.  The ambiguity introduced by the two final-state neutrinos was resolved using the so-called neutrino weighting method\footnote{Refer to the following publication by the D0 collaboration \cite{NWM}.} in order to reconstruct a \ttbar \ system for each candidate event.  Both the absolute and normalized measured cross sections were presented.  Overall good agreement was observed except in the case of one Powheg-Box+Herwig++ prediction where moderate disagreement was seen in the $\pT^{t}$ and $\mass{}^{\ttbar}$ distributions.  Similar measurements in the dileptonic channel were performed by CMS \cite{CMS29}.

Differential \ttbar \ cross-section measurements were also performed in the $\ell$+jets decay channel by both ATLAS \cite{ATLAS7} and CMS \cite{CMS30,CMS31}.  In the case of the ATLAS measurement, based on 3.2 \ifb \ of $\rts=13$ \TeV \ data, the resolved and boosted \ttbar \ regimes are targeted separately, each with their own dedicated object and event selection.  As in the case of the dileptonic channel, the differential cross sections were presented both as absolute and normalized fiducial measurements, and for a number of similar kinematic observables.  The general observation was that the spectra based on NLO predictions were harder than those observed in the measured data.  This was seen to be present in both the boosted and resolved results.  The trend was further observed in a separate dedicated and recently published analysis, also by the ATLAS collaboration, of differential cross-section measurements of \ttbar \ production in the $\ell$+jets channel, but separated into exclusive $N_{\rm jet}$ bins \cite{ATLAS8}.  The final normalized distribution as a function of the transverse momentum of the hadronic top quark candidate ($\pT^{t,{\rm had}}$) is shown in Figure~\ref{Fig-ATLAS-8} specifically for the $N_{\rm jet}=4$ case.  The general observation remains the same as in the inclusive measurement, and the effect at high $\pT^{t,{\rm had}}$ is most pronounced at lower jet multiplicity ($N_{\rm jet}=4$), but is also present for the higher jet multiplicities.

\begin{figure}[!htbp]
\begin{subfigure}{0.47\textwidth}
\begin{centering}
\includegraphics[scale=0.35]{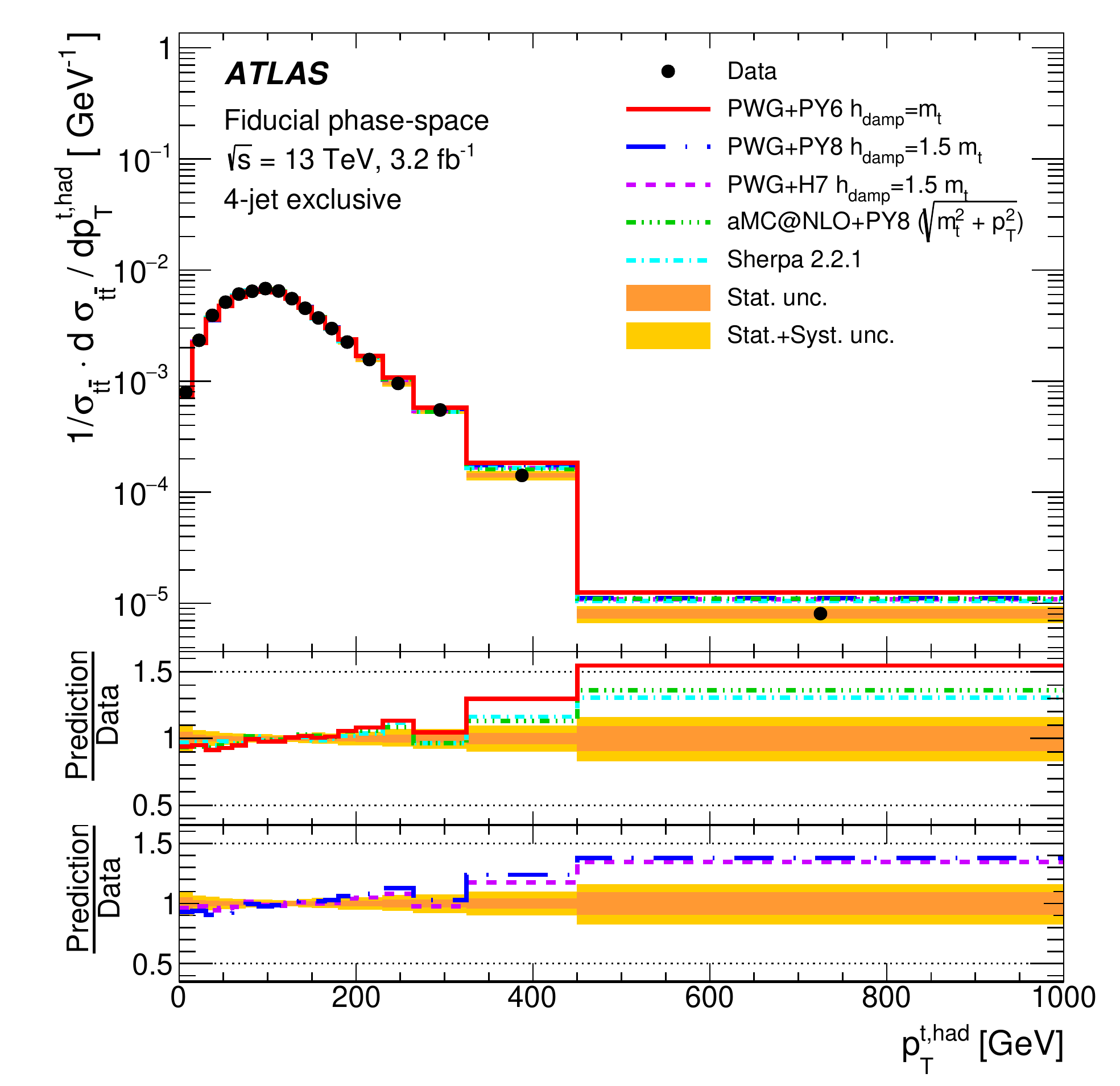}
\caption{}
\label{Fig-ATLAS-8}
\end{centering}
\end{subfigure} 
\begin{subfigure}{0.47\textwidth}
\begin{centering}
\includegraphics[scale=0.35]{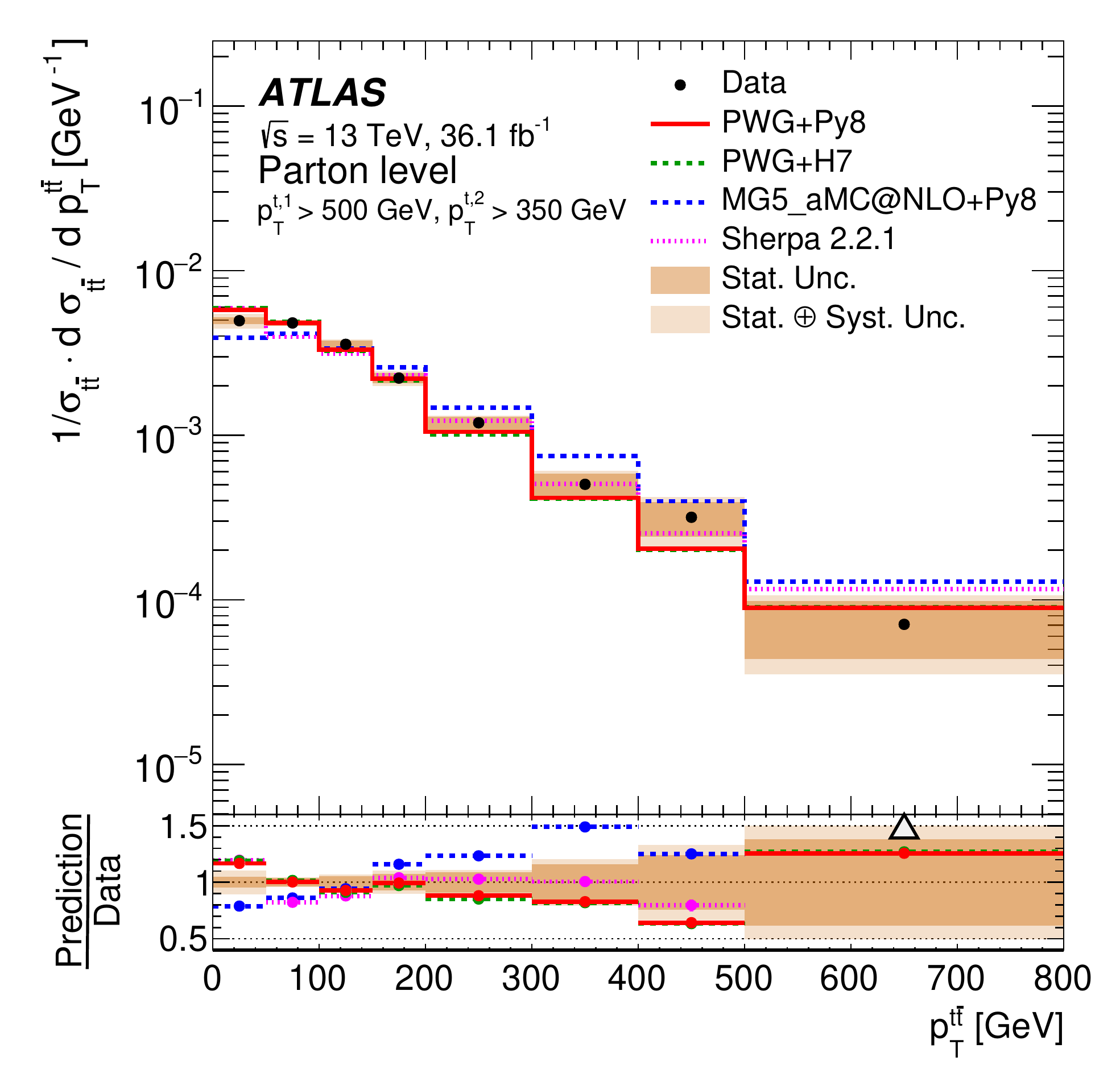}
\caption{}
\label{Fig-ATLAS-9}
\end{centering}
\end{subfigure} 
\label{Fig-ATLAS-8-ATLAS-9}
\caption{The measured normalized differential \ttbar \ cross-sections, as measured by ATLAS, as a function of (a) the hadronic top quark \pT \ spectrum in the exclusive $N_{\rm jet}=4$ bin for the $\ell$+jets analysis \cite{ATLAS8} and (b) the \pT \ of the reconstructed \ttbar \ system in the boosted all-hadronic analysis \cite{ATLAS9}.}
\end{figure}

 
A dedicated \ttbar \ differential cross-section measurement was performed by the ATLAS collaboration in the all-hadronic channel by searching for events compatible with a boosted \ttbar \ final state \cite{ATLAS9}.  Advantages of the all-hadronic \ttbar \ channel include a high branching ratio and no neutrinos in the hard-scatter final state, allowing for a full event reconstruction.  In contrast, the lack of an isolated high-\pT \ lepton in the final state leads to a formidable source of background from multijet events which must be estimated using a data-driven approach.  The analysis targets events featuring two large-radius ($R=1.0$) jets and employs so-called jet grooming and top-tagging techniques to increase the relative fraction of signal events.  Corrections for detector-level effects, efficiencies and overall acceptance led to a final set of distributions unfolded to particle and parton level within the fiducial volume.  Figure~\ref{Fig-ATLAS-9} shows the resulting normalized differential cross section as a function of the transverse momentum of the \ttbar \ system, showing good overall agreement between the measured data and simulation.  The notable exception is a slight disagreement in the MadGraph aMC@NLO prediction interfaced with Pythia8 for the parton shower and hadronization modelling; in this case the simulation predicts a harder $\pT^{\ttbar}$ spectrum compared with what is observed in the data.  The overall event yields were seen to be somewhat lower in data relative to the NLO predictions -- though this is recognizably not visible in the figure shown -- but are nevertheless consistent within the overall uncertainty bands.


\subsection{Non-Standard $\ttbar$ Production Results}

Production cross-section measurements for \ttbar \ pairs at the LHC, including those mentioned above, have primarily focused on $p$-$p$ collision data collected at the standard centre-of-mass energies $\rts =$ 7, 8, and 13 \TeV.  The CMS collaboration has recently published two complementary such measurements.  The first was based on the small 27.4 \ipb \ $p$-$p$ dataset collected at $\rts = 5.02$ \TeV \ \cite{CMS34}.  In addition to achieving a relative precision on $\sigma_{\ttbar}$ of roughly 12\% and the inclusion of an additional data point in summary plots in order to compare the data with predictions for the dependence of $\sigma_{\ttbar}$ on $\rts$, the measurement allows for a moderate improvement of the gluon PDF uncertainty at high $x$. The second, non-standard measurement was based on $p$-$Pb$ collisions at a nucleon-nucleon centre-of-mass energy of $\sqrt{s_{\rm NN}} = 8.16$ \TeV \ \cite{CMS35}.  The clear presence of a signal peak can be seen in Figure~\ref{Fig-CMS-35}, which shows the signal-dominated distribution of the invariant mass of reconstructed top quark candidates in one of the final signal regions for this analysis, where a single lepton ($e/\mu$) and at least two b-tagged jets plus two additional jets were required.  The measured $p$-$Pb$ cross-section result was scaled in order to be able to compare it to $p$-$p$ measurements at $\rts = 8$ \TeV.  Any discrepancies between the scaled results -- though none were observed -- could potentially shed light into the way gluons are distributed in high-$Z$ nuclei. 


\begin{figure}[!htbp]
\begin{subfigure}{0.47\textwidth}
\begin{centering}
\includegraphics[scale=0.32]{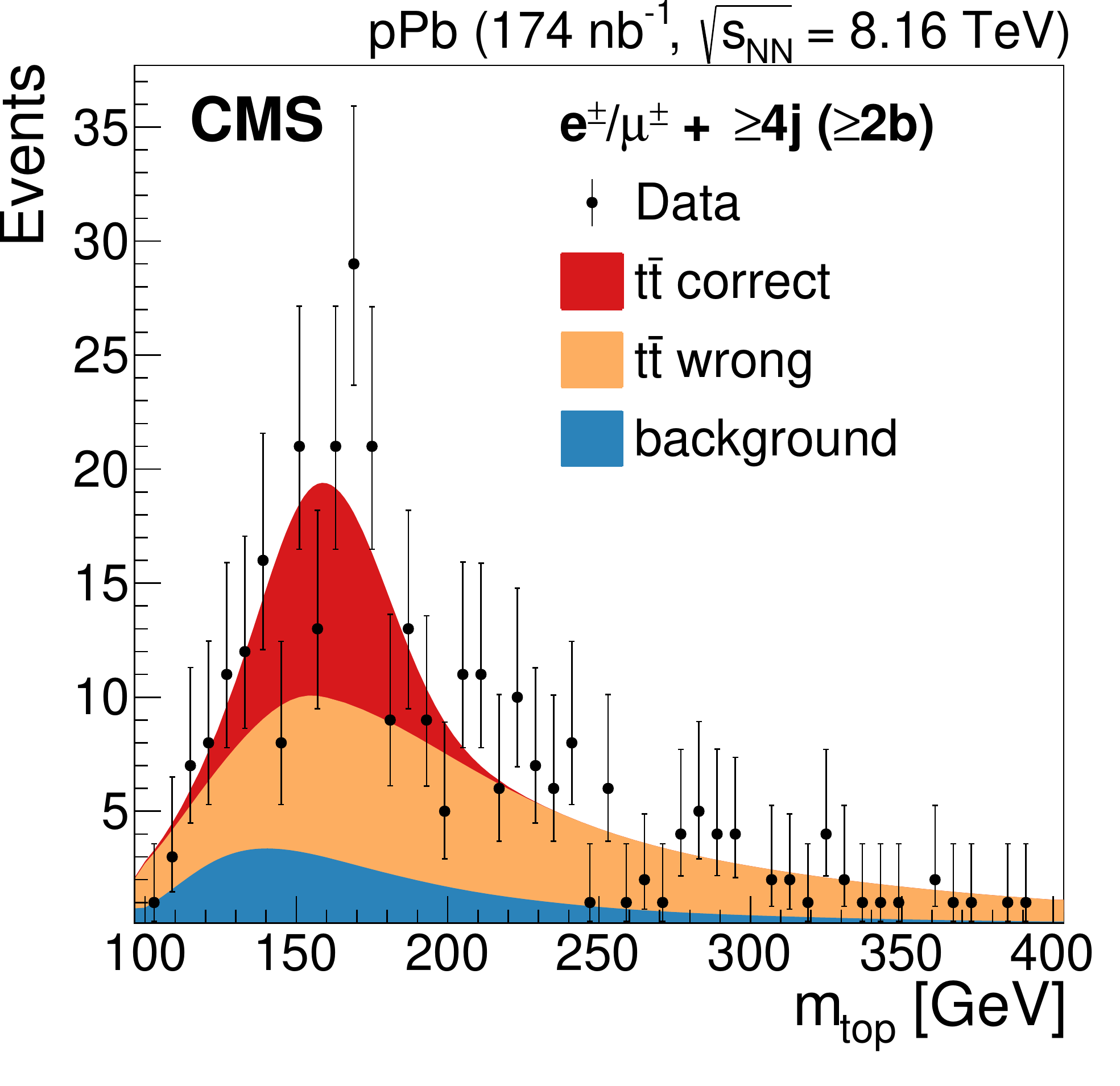}
\caption{}
\label{Fig-CMS-35}
\end{centering}
\end{subfigure} 
\begin{subfigure}{0.47\textwidth}
\begin{centering}
\includegraphics[scale=0.43]{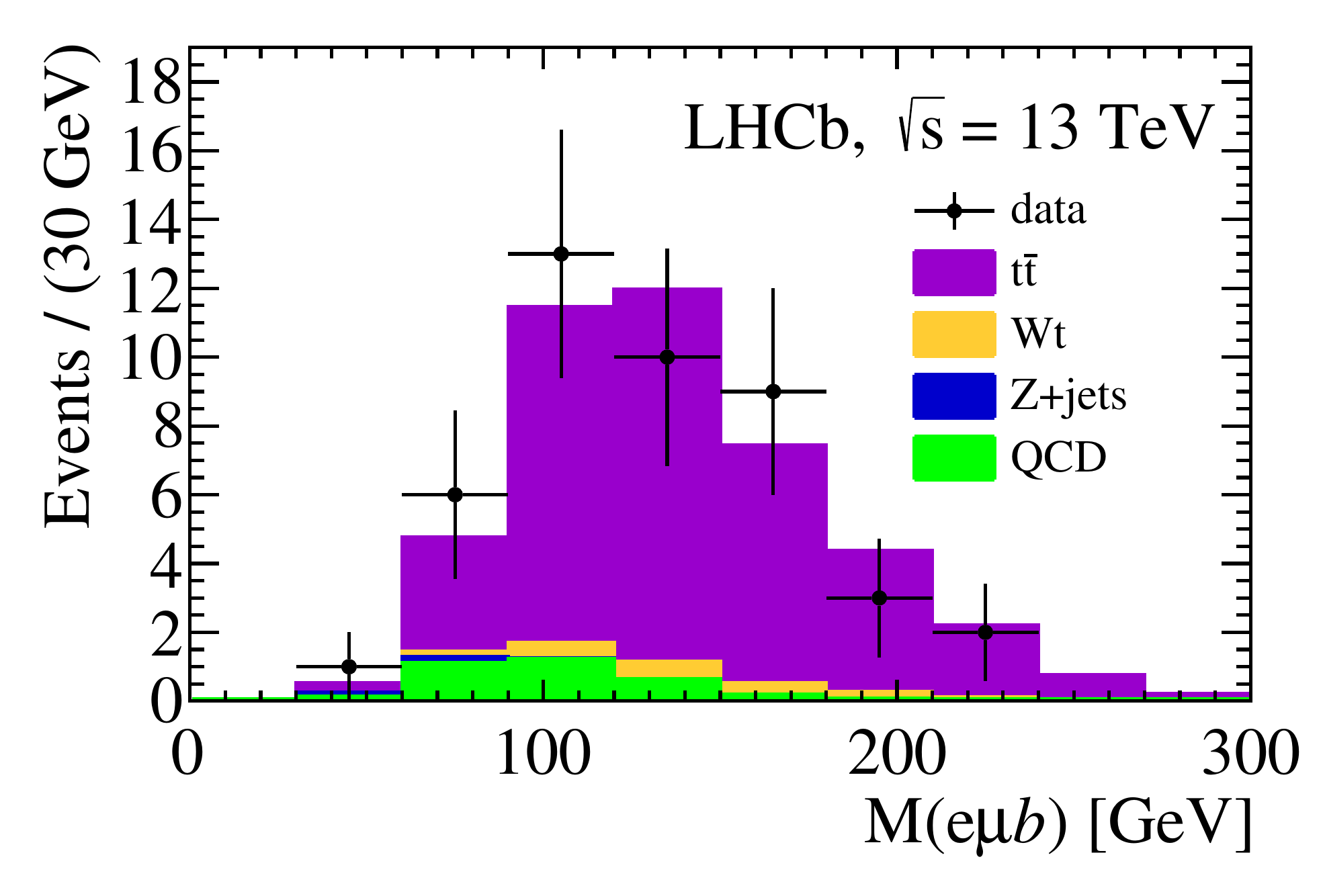}
\caption{}
\label{Fig-LHCb-46}
\end{centering}
\end{subfigure} 
\label{Fig-CMS-35-LHCb-46}
\caption{Distributions of (a) the reconstructed top quark mass of all final candidates in the \ttbar \ cross-section measurement performed at $\rts=5.02$ \TeV \  by CMS \cite{CMS34} and (b) the invariant $e\mu b$ mass for candidate top quark events in a recent LHCb analysis \cite{LHCb46}.}
\end{figure}


Although LHC results featuring top quarks are typically limited to ATLAS \& CMS analyses, a recent LHCb publication details a search for \ttbar \ pairs produced in the high-$\eta$ region of the detector \cite{LHCb46}.  The result highlights a substantial increase in the fraction of signal top quark events (Figure~\ref{Fig-LHCb-46}) compared to previous such background-dominated searches using the Run 1 datasets \cite{LHCb44,LHCb45}, and ushers in a new era of top quark measurements by LHCb to complement those by ATLAS and CMS. The $\qqbar $ production modes play a larger relative role in forward-produced \ttbar \ pairs compared with $gg$ modes, which could allow for heightened sensitivity to the \ttbar \ charge asymmetry -- an effect driven by higher-order $\qqbar$-mode production diagrams.  In addition, such measurements offer improved sensitivity to the higher-$x$ regions of the gluon PDF.  Present LHCb top quark measurements, including this analysis, are recognizably limited by statistics, and the lower overall acceptance precludes a full event reconstruction or a realiable measurement of \met.   The analysis nevertheless demonstrates that the future of the top quark physics programme at LHCb is indeed very promising, as is the prospect of the eventual inclusion of its results in future combinations with the other LHC experiments.


\subsection{Rare Searches Involving Top Quarks}

Searches for the associated production of a \ttbar \ pair and a vector boson via the process $pp\to\ttbar V$  (where $V=\Wboson,\Zboson$) were carried out at $\rts = 13$ \TeV \ by both the ATLAS and CMS collaborations.  Although previous analyses have been performed which directly probe the Higgs boson couplings to vector bosons via $\Wboson\Wboson^{*}$, $\Zboson\Zboson^{*}$, $\gamma\gamma$, or $\tau\tau$, direct probes of the SM $\Hboson qq$ coupling have until recently remained beyond the possible experimental reach.  The observation and further studies of \ttbar$V$ production in particular allows the Higgs boson coupling to the top quark to be better understood.  The CMS search was performed in the same-sign dilepton ($\ell^{\pm}\ell^{\pm}$)+jets channel for the case of \ttbar\Wboson, and in the 3$\ell$- and 4$\ell$+jets channels for the case of \ttbar\Zboson \ \cite{CMS36}.  The integrated luminosity of the considered dataset corresponded to 35.9 \ifb. The $3\ell$ channel offered the greatest sensitivity; the event yields for the final signal regions associated with this decay channel can be seen in Figure~\ref{Fig-CMS-36}, where good agreement can be observed between the measured data and SM predictions.  The various bins correspond to the combinations of reconstructed lepton flavours, and one can note that in all subchannels the signal process is dominant.

The \ttbar\Wboson \ and \ttbar\Zboson \ cross sections were extracted simultaneously, resulting in an observed significance of 4.8$\sigma$ for \ttbar\Wboson and larger than 5$\sigma$ for \ttbar\Zboson.  The ATLAS result based on a smaller 3.2 \ifb \ dataset also considers several distinct signal regions, and yields an observed signifance of 2.2$\sigma$ and 3.9$\sigma$ for the $\ttbar\Wboson$ and $\ttbar\Zboson$ processes, respectively \cite{ATLAS10}.  The details of the ATLAS measurement are presented separately in a dedicated contribution.

\begin{figure}[!htbp]
\begin{subfigure}{0.47\textwidth}
\begin{centering}
\includegraphics[scale=0.32]{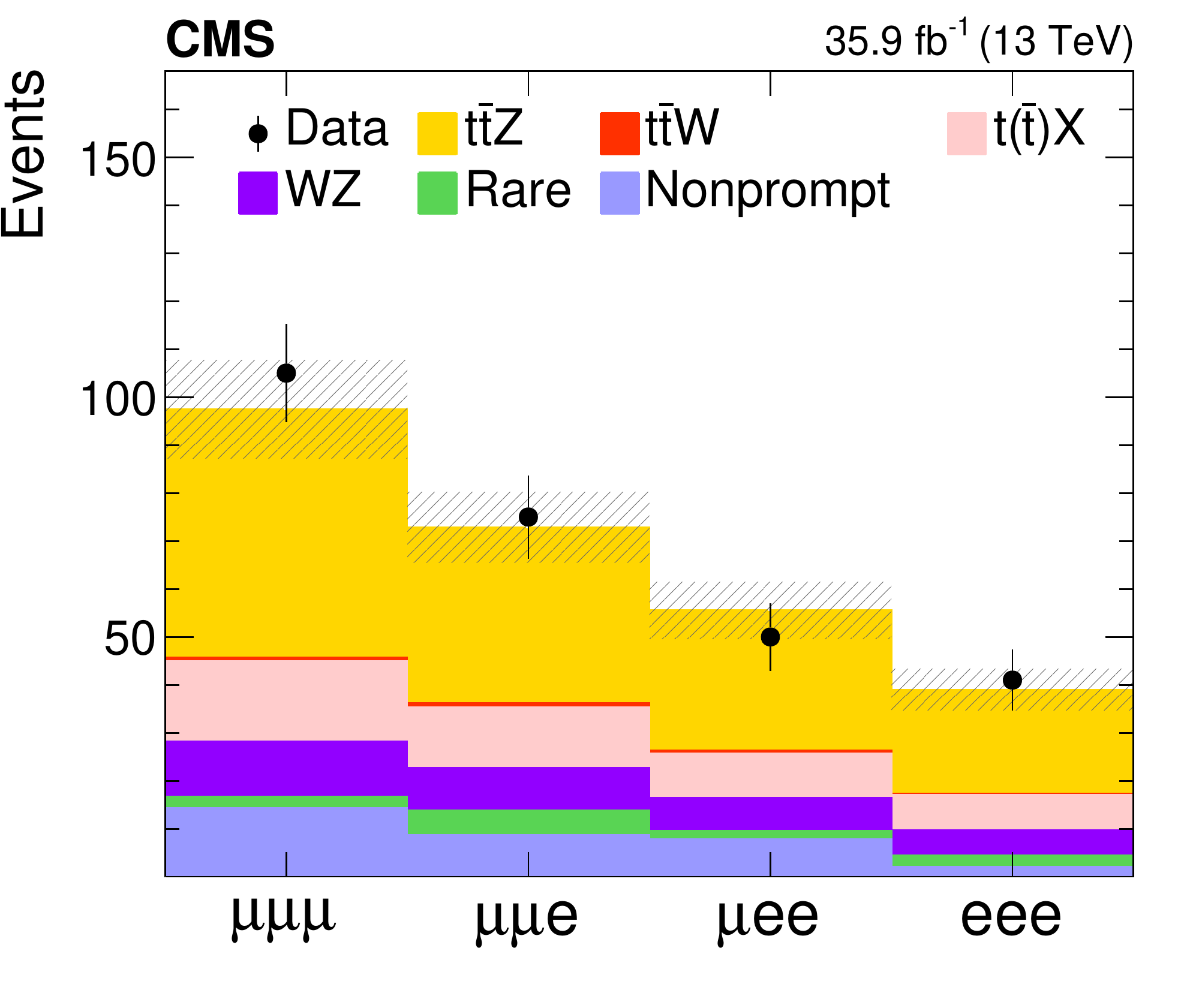}
\caption{}
\label{Fig-CMS-36}
\end{centering}
\end{subfigure} 
\begin{subfigure}{0.47\textwidth}
\begin{centering}
\includegraphics[scale=0.39]{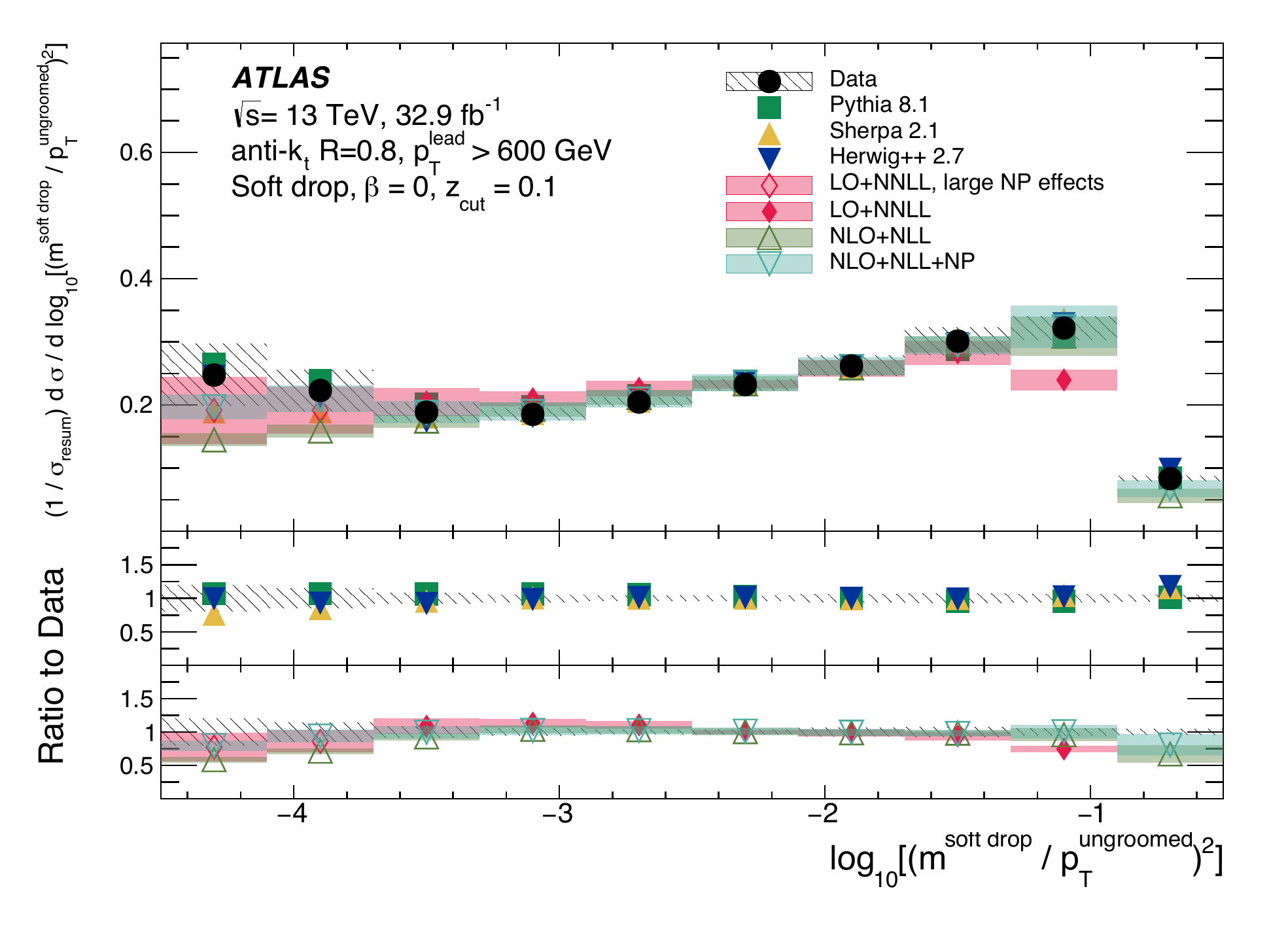}
\caption{}
\label{Fig-ATLAS-17}
\end{centering}
\end{subfigure} 
\label{Fig-CMS-36-ATLAS-17}
\caption{(a) Final event yields in the $3\ell$ decay channel for the $\ttbar\Zboson$ cross-section measurement performed by CMS \cite{CMS36} and (b) one of the measured normalized distributions from the jet soft-drop analysis based on dijet events at $\rts=13$ \TeV \ from ATLAS \cite{ATLAS17}.}
\end{figure}


The combination of a higher centre-of-mass energy and the increased statistics collected in recent years have put a search for four-top production (\ttbar\ttbar) within experimental reach at the LHC.  Dedicated searches for this rare SM process have recently been carried out by both the ATLAS and CMS collaborations.  The ATLAS result employs 3.2 \ifb \ of integrated luminosity and a base event selection requiring a single reconstructed lepton ($e/\mu$) \cite{ATLAS14}.  Candidate events were separated into a number of dedicated control (6), validation (9) and signal (3) regions based on the reconstructed jet and b-jet multiplicities. The largest signal-to-background ratio ($S/B\approx 4.5\%$) was obtained in a region requiring at least 10 hadronic jets, of which at least 4 must be identified by a dedicated b-tagging algorithm.  The final result is an observed (expected) upper limit of 21 (18) times the SM prediction for $\sigma_{\ttbar\ttbar}$ at the 95\% C.L.

Similar searches for \ttbar\ttbar \ production were carried out by the CMS collaboration.  An initial result based on 2.6 \ifb \ of Run 2 data was recently superseded by an extended analysis employing 35.6 \ifb of data and requiring at least two same-sign leptons \cite{CMS39,CMS40}.  As in the case of the ATLAS analysis, the final set of candidate events were separated into various control and signal regions, in this case based on jet, b-jet and lepton multiplicities, as well as the value of the reconstructed invariant dilepton mass ($\mass{\ell}$).  A slight excess of candidate \ttbar\ttbar \ events at the 1.6$\sigma$ level was observed over a background-only hypothesis, and a value of $\sigma_{\ttbar\ttbar} = 16.9^{+13.8}_{-11.4}$ pb was extracted.  Moreover, the presence of virtual Higgs boson loops in certain \ttbar\ttbar \ production diagrams allowed $\sigma_{\ttbar\ttbar}$ to be expressed as a function of the top quark Yukawa coupling relative to its SM value ($|y_{t}/y_{t}^{\rm SM}|$), thereby allowing constraints to be placed on this ratio based on the measured result.  

It is interesting to note that in such \ttbar\ttbar \ searches the processes $\ttbar\Hboson$ and $\ttbar V$ (where $V=\Wboson,\Zboson$) -- rare SM processes themselves and only recently able to be measured -- constitute a considerable fraction of the background.  The outlook for such four-top searches with larger LHC datasets in the near future is promising.

\noindent\rule[0.5ex]{\linewidth}{0.5pt}
\section{Recent Results Involving Boosted Objects and Jet Substructure}

One of the key motivations for studies of jet substructure is in providing enhanced discrimination between signal and backgrounds for a variety of analyses.  Such discrimination can be at the single-object level, such as the ability to separate quark- from gluon-initiated jets, where the latter constitute a substantial source of QCD background in a $p$-$p$ collision environment, but it can also prove useful in identifying cases in which the decay products of a boosted, heavy SM particle (e.g. $\Wboson$, $\Zboson$, $\Hboson$, or $t$) are highly collimated, leading to the reconstruction of a single, large-R jet.  Recent advances in jet grooming and multivariate analysis techniques, but also in analytical methods in understanding the effects of small- or wide-angle gluon radiation, have offered additional benefits.  Since one of the possible telltale signatures of new, heavy BSM models is the presence of boosted SM particles, advances in methods exploiting knowledge of jet substructure will lead to an improved sensitivity to new physics.


Recent developments in analytical methods have introduced jet substructure observables accurate beyond leading-logarithm accuracy.  One recent ATLAS publication based on 32.9 \ifb \ of $\rts=13$ \TeV \ data presents a first measurement involving the jet soft-drop mass -- one such substructure variable -- using dijet events from $p$-$p$ collisions \cite{ATLAS17}.  The analysis features a re-clustering of the constituents of $\akt$ jets based on a designated soft-drop mass procedure.  This procedure is based on an iterative condition which is governed primarily by the choice of two parameters: $z_{\rm cut}$ and $\beta$, which relate to $\pT$ and angular thresholds in the context of gluon radiation.  The final results correspond to normalized differential cross-section measurements, presented as a function of the logarithm of the dimensionless quantity $\rho^2$, where $\rho$ is the ratio of the invariant mass of the jet following the soft-drop procedure and the transverse momentum of the ungroomed jet ($\rho = \mass{}^{\rm soft \ drop}/\pT^{\rm ungroomed}$).  The inclusion of $\pT^{\rm ungroomed}$ in the denominator counters an undesireable $\pT$ dependence.  The distribution shown in Figure~\ref{Fig-ATLAS-17}, corresponding to the particular choice of $\beta=0$, can be contrasted with other such distributions (for $\beta=1,2$) in the original publication.  An unfolding procedure is used to correct for detector-level effects. The evolution of the distributions with $\beta$ and $z_{\rm cut}$ highlights particular regions of the quantity $\log_{10}\rho^2$ with larger levels of disagreement between data and predictions. Such disagreements between the measured data and either precise QCD calculations or leading-logorithm Monte Carlo simulations will prove beneficial in improving our understanding of the collinear QCD regime. 


Dedicated jet substructure studies were performed in a very recent measurement by the CMS collaboration which targeted \ttbar \ events in the $\ell$+jet ($\ell=e,\mu$) decay channel \cite{CMS32}, complementing the earlier-mentioned \ttbar \ cross-section measurements.  Such an analysis probes the perturbative vs. non-perturbative phases of jet evolution.  Results were presented both inclusively in terms of jet flavour, as well as separately for bottom-quark-, light-quark-, and gluon-initiated jets.  The final distributions were presented as a function of a number of observables, including one category encompassed by the generalized variable $\lambda_{\beta}^{\kappa} = \sum_{i}z_{i}^{\kappa}\left(\frac{\Delta R(i,\hat{n}_{r})}{R}\right)^{\beta}$, governed by parameters similar to $z_{\rm cut}$ and $\beta$ mentioned in the ATLAS soft-drop mass analysis above\footnote{Explicit terms in the above expression are: the \pT \ fraction carried by constituent $i$ ($z_{i} = p_{T,i}/\sum_{j}p_{T,j})$; the distance parameter of the jet algorithm (R); and the separation in $\eta$-$\phi$ space between constituent $i$ and the so-called recoil-free jet axis ($\Delta R(i,\hat{n}_{r})$).  For additional information refer to the CMS document.}.  The sum in the expression is taken over jet constituents and leads to several particular examples for choices of $\kappa$ and $\beta$: the number of jet constituents ($\lambda_{0}^{0}$), the jet width ($\lambda_{1}^{1}$), and the jet \pT \ dispersion ($\lambda_{2}^{0}$).  The results allow for refinements via the tuning of simulations. Additionally the distribution for the jet width observable allowed for an extraction of the effective value of the strong coupling constant ($\alphas^{\rm FSR}$). Yet another recent CMS result explored a study of the underlying event in \ttbar \ events \cite{CMS33}.



A search was performed by the ATLAS collaboration for new resonances decaying via two heavy gauge bosons ($VV$, where $V=\Wboson,\Zboson$) \cite{ATLAS18}.  The $\rts=13$ \TeV \ result based on a 36.7 \ifb \ dataset employed a search strategy which targeted hadronically decaying high-\pT \ $\Wboson/\Zboson$ bosons.  Initially formed large-R jets were reclustered into smaller $R=0.2$ jets and trimming techniques were applied.  A number of substructure variables were used to heighten the sensitivity to true $VV$ processes.  The mass of selected jet candidates proves to be a highly discriminating observable, but can be measured either with track- or calorimeter-based information; the results from both methods were combined to yield a single mass value, with a weight corresponding to their respective mass resolutions based on simulation.
Several final signal regions ($\Wboson\Wboson$, $\Wboson\Zboson$, $\Zboson\Zboson$) were ultimately considered, and the results provide sensitivity to new resonances in the 1.2 - 5.0 TeV mass range.  No excesses above  SM predictions were observed, and several limits were placed at the 95\% C.L. on the cross section times branching ratio for given mass hypotheses.


Another search for new heavy $V\Hboson$ resonances in the $V\to\qqbar,\Hboson\to\bbbar$ channel was performed by the ATLAS collaboration in an analysis requiring two boosted, large-R jets \cite{ATLAS19}\footnote{This was complemented by a similar recent ATLAS measurement searching for heavy $V\Hboson$ resonances decaying via $\Hboson\to\bbbar$ and either $\Zboson\to\ell\ell,\nu\nu$ or $\Wboson\to\lnu$ \cite{ATLAS20}.}. The dominant multijet background, making up roughly 90\% of the final event composition, was estimated using a data-driven technique.  The number of b-tagged jets, as well as the values of the invariant $\qqbar$ and $\bbbar$ masses, were used to classify the set of selected events into signal, sideband and validation regions.  Jet substructure variables, as well as the combined jet mass described for the $VV$ analysis mentioned above, were used to enhance the signal significance. Overall a good agreement between data and SM predictions was observed, with the largest excess seen at an invariant dijet mass ($\mass{\rm J}$) of 3.0 TeV, corresponding to a local (global) significance of 3.3 (2.1) $\sigma$.  The results of the search allowed constraints to be placed on heavy vector triplets models.

\noindent\rule[0.5ex]{\linewidth}{0.5pt}
\section{Summary}

The excellent performance of the ATLAS, CMS and LHCb experiments at the LHC in recent years has allowed for great strides to be made in our understanding of the physics of the SM.  Several extremely rare SM processes, often featuring multiple heavy SM particles in the hard-scatter final state, have recently been observed. The outlook for extended results based on the full Run 2 datasets in all three experiments is promising.  No significant deviations from SM predictions have been observed so far.  One prominant change in recent months is the inclusion of LHCb as a source for top-quark physics results which will complement the top quark programmes of ATLAS and CMS.  The development of techniques to identify boosted SM objects via substructure and multivariate-based methods is a driving force for pushing the limits in sensitivity in many recent searches; their role in future analyses will be of the utmost importance.  The analyses presented here recognizably represent but a small subset of exciting new results from the LHC experiments.


\begin{thebibliography}{99}


\bibitem{ATLASExperiment}
ATLAS Collaboration, \emph{The ATLAS Experiment at the CERN Large Hadron Collider}, \\ JINST {\bf 3} (2008) S08003 [\href{http://inspirehep.net/record/796888}{\tt inspirehep.net/record/796888}].

\bibitem{CMSExperiment}
CMS Collaboration, \emph{The CMS Experiment at the CERN LHC}, JINST {\bf 3} (2008) S08004 [\href{http://inspirehep.net/record/796887}{\tt inspirehep.net/record/796887}].

\bibitem{LHCbExperiment}
LHCb Collaboration, \emph{The LHCb Detector at the LHC}, JINST {\bf 3} (2008) S08005 [\href{http://inspirehep.net/record/796248}{\tt inspirehep.net/record/796248}].


\bibitem{ATLAS1}
ATLAS Collaboration, \emph{Measurement of inclusive jet and dijet cross-sections in proton-proton collisions at $\sqrt{s}$ = 13 TeV with the ATLAS detector}, JHEP {\bf 05} (2018) 195 [\href{https://arxiv.org/abs/1711.02692}{\tt arXiv:1711.02692}].

\bibitem{antiktalgo}
M. Cacciari, G. P. Salam, and G. Soyez, \emph{The anti-$k_{t}$ jet clustering algorithm}, JHEP {\bf 04} (2008) 063 [\href{https://arxiv.org/abs/0802.1189}{\tt arXiv:0802.1189v2}].

\bibitem{CMS21}
CMS Collaboration, \emph{Measurement of the triple-differential dijet cross section in proton-proton collisions at $\sqrt{s}$ = 8 TeV and constraints on parton distribution functions}, \\ Eur. Phys. J. C {\bf 77} (2017) 746 [\href{https://arxiv.org/abs/1705.02628}{\tt arXiv:1705.02628v2}].

\bibitem{CMS22}
CMS Collaboration, \emph{Measurement of the double-differential inclusive jet cross section in proton-proton collisions at $\sqrt{s}$ = 13 TeV}, Eur. Phys. J. C {\bf 76} (2016) 451 [\href{https://arxiv.org/abs/1605.04436}{\tt arXiv:1605.04436v2}].

\bibitem{CMS23}
CMS Collaboration, \emph{Measurement and QCD analysis of double-differential inclusive jet cross sections in pp collisions at $\sqrt{s}$ = 8 TeV and ratios to 2.76 and 7 TeV}, JHEP {\bf 03} (2017) 156 [\href{https://arxiv.org/abs/1609.05331}{\tt arXiv:1609.05331v2}].

\bibitem{CMS28}
CMS Collaboration, \emph{Measurement of the differential cross sections for the associated production of a \Wboson \ boson and jets in proton-proton collisions at $\sqrt{s}$ = 13 TeV}, Phys. Rev. D {\bf 96} (2017) 072005 [\href{https://arxiv.org/abs/1707.05979}{\tt arXiv:1707.05979v2}].

\bibitem{CMS24}
CMS Collaboration, \emph{Observation of electroweak production of same-sign \Wboson \ boson pairs in the two jet and two same-sign lepton final state in proton-proton collisions at $\sqrt{s}$ = 13 TeV}, \\ Phys. Rev. Lett. {\bf 120} (2018) 081801 [\href{https://arxiv.org/abs/1709.05822}{\tt arXiv:1709.05822v2}].

\bibitem{ATLAS5}
ATLAS Collaboration, \emph{Measurement of the $\Wplus\Wminus$ production cross section in pp collisions at a centre-of-mass energy of $\sqrt{s}$ = 13 TeV with the ATLAS experiment}, Phys. Lett. B {\bf 773} (2017) 354 [\href{https://arxiv.org/abs/1702.04519}{\tt arXiv:1702.04519v2}].

\bibitem{LHCb43}
LHCb Collaboration, \emph{First observation of forward $\Zboson\to\bbbar$ production in pp collisions at $\sqrt{s}$ = 8 TeV}, Phys. Lett. B {\bf 776} (2017) 430-439 [\href{https://arxiv.org/abs/1709.03458}{\tt arXiv:1709.03458v2}].

\bibitem{CMS25}
CMS Collaboration, \emph{Electroweak production of two jets in association with a \Zboson \ boson in proton-proton collisions at $\sqrt{s}$ = 13 TeV}, Submitted to Eur. Phys. J. C (December 28, 2017) [\href{https://arxiv.org/abs/1712.09814}{\tt arXiv:1712.09814}].

\bibitem{CMS26}
CMS Collaboration, \emph{Measurements of the pp$\to$\Zboson\Zboson \ production cross section and the Z$\rightarrow$4$\ell$ branching fraction, and constraints on anomalous triple gauge couplings at $\sqrt{s}$ = 13 TeV}, \\ Eur. Phys. J. C {\bf 78} (2018) 165 [\href{https://arxiv.org/abs/1709.08601}{\tt arXiv:1709.08601v2}].

\bibitem{CMS27}
CMS Collaboration, \emph{Measurement of vector boson scattering and constraints on anomalous quartic couplings from events with four leptons and two jets in proton-proton collisions at $\sqrt{s}$ = 13 TeV}, \\ Phys. Lett. B {\bf 774} (2017) 682 [\href{https://arxiv.org/abs/1708.02812}{\tt arXiv:1708.02812v2}].

\bibitem{CMSPFlowJets}
CMS Collaboration, \emph{Particle-flow reconstruction and global event description with the CMS detector}, JINST {\bf 12} (2017) P10003 [\href{https://arxiv.org/abs/1706.04965}{\tt arXiv:1706.04965v2}].

\bibitem{ATLAS2}
ATLAS Collaboration, \emph{Measurements of the production cross section of a \Zboson \ boson in association with jets in pp collisions at $\sqrt{s}$ = 13 with the ATLAS detector}, Eur. Phys. J. C {\bf 77} (2017) 361 [\href{https://arxiv.org/abs/1702.05725}{\tt arXiv:1702.05725v2}].

\bibitem{ATLAS3}
ATLAS Collaboration, \emph{Measurement of the cross-section for electroweak production of dijets in association with a \Zboson \ boson in pp collisions at $\sqrt{s}$ = 13 TeV with the ATLAS detector}, \\ Phys. Lett. B {\bf 775} (2017) 206 [\href{https://arxiv.org/abs/1709.10264}{\tt arXiv:1709.10264v3}].

\bibitem{ATLAS4}
ATLAS Collaboration, \emph{ZZ$\rightarrow\ell^{+}\ell^{-}\ell^{\prime +}\ell^{\prime -}$ cross-section measurements and search for anomalous triple gauge couplings in 13 TeV pp collisions with the ATLAS detector}, Phys. Rev. D {\bf 97} (2018) 032005 [\href{https://arxiv.org/abs/1709.07703}{\tt arXiv:1709.07703v2}].


\bibitem{ATLAS6}
ATLAS Collaboration, \emph{Measurements of top-quark pair differential cross-sections in the $e\mu$ channel in pp collisions at $\sqrt{s}$ = 13 TeV using the ATLAS detector}, Eur. Phys. J. C {\bf 77} (2017) 299 [\href{https://arxiv.org/abs/1612.05220}{\tt arXiv:1612.05220v2}].

\bibitem{NWM}
D0 Collaboration, \emph{Measurement of the Top Quark Mass Using Dilepton Events}, \\ Phys. Rev. Lett. {\bf 80} (1998) 2063-2068 [\href{https://arxiv.org/abs/hep-ex/9706014}{\tt arXiv:hep-ex/9706014}].

\bibitem{CMS29}
CMS Collaboration, \emph{Measurement of normalized differential \ttbar \ cross sections in the dilepton channel from pp collisions at $\sqrt{s}$ = 13 TeV}, JHEP {\bf 04} (2018) 060 [\href{https://arxiv.org/abs/1708.07638}{\tt arXiv:1708.07638v2}].

\bibitem{ATLAS7}
ATLAS Collaboration, \emph{Measurements of top-quark pair differential cross-sections in the lepton+jets channel in pp collisions at $\sqrt{s}$ = 13 TeV using the ATLAS detector}, JHEP {\bf 11} (2017) 191 [\href{https://arxiv.org/abs/1708.00727}{\tt arXiv:1708.00727v2}].

\bibitem{CMS30}
CMS Collaboration, \emph{Measurement of differential cross sections for the production of top quark pairs and of additional jets in lepton+jets events from pp collisions at $\sqrt{s}$ = 13 TeV}, \\ Submitted to Phys. Rev. D (March 23, 2018) [\href{https://arxiv.org/abs/1803.08856}{\tt CMS-PAS-TOP-17-002}].

\bibitem{CMS31}
CMS Collaboration, \emph{Measurements of differential cross sections of top quark pair production as a function of kinematic event variables in proton-proton collisions at $\sqrt{s}$ = 13 TeV}, JHEP {\bf 06} (2018) 002 [\href{https://arxiv.org/abs/1803.03991}{\tt arXiv:1803.03991v2}].

\bibitem{ATLAS8}
ATLAS Collaboration, \emph{Measurements of differential cross sections of top quark pair production in association with jets in pp collisions at $\sqrt{s}$ =13 TeV using the ATLAS detector}, Submitted to JHEP (February 19, 2018) [\href{https://arxiv.org/abs/1802.06572}{\tt arXiv:1802.06572}].

\bibitem{ATLAS9}
ATLAS Collaboration, \emph{Measurements of \ttbar \ differential cross-sections of highly boosted top quarks decaying to all-hadronic final states in pp collisions at $\sqrt{s}$ = 13 TeV using the ATLAS detector}, Submitted to Phys. Rev. D (January 6, 2018) [\href{https://arxiv.org/abs/1801.02052}{\tt arXiv:1801.02052}].

\bibitem{CMS34}
CMS Collaboration, \emph{Measurement of the inclusive \ttbar \ cross section in pp collisions at $\sqrt{s}$ = 5.02 TeV using final states with at least one charged lepton}, JHEP {\bf 03} (2018) 115 [\href{https://arxiv.org/abs/1711.03143}{\tt arXiv:1711.03143v2}].

\bibitem{CMS35}
CMS Collaboration, \emph{Observation of top quark production in proton-nucleus collisions}, \\ Phys. Rev. Lett. {\bf 119} (2017) 242001 [\href{https://arxiv.org/abs/1709.07411}{\tt arXiv:1709.07411v2}].

\bibitem{LHCb46}
LHCb Collaboration, \emph{Measurement of forward top pair production in the dilepton channel in pp collisions at $\sqrt{s}$ = 13 TeV}, Submitted to JHEP (March 14, 2018) [\href{https://arxiv.org/abs/1803.05188}{\tt arXiv:1803.05188}].

\bibitem{LHCb44}
LHCb Collaboration, \emph{First observation of top quark production in the forward region}, \\ Phys. Rev. Lett. {\bf 115} (2015) 112001 [\href{https://arxiv.org/abs/1506.00903}{\tt arXiv:1506.00903v2}].

\bibitem{LHCb45}
LHCb Collaboration, \emph{Measurement of forward $\ttbar$, $\Wboson+\bbbar$, $\Wboson+\ccbar$ production in pp collisions at \\ $\sqrt{s}$ = 8 TeV}, Phys. Lett. B {\bf 767} (2017) 110 [\href{https://arxiv.org/abs/1610.08142}{\tt arXiv:1610.08142v2}].

\bibitem{CMS36}
CMS Collaboration, \emph{Measurement of the cross section for top quark pair production in association with a \Wboson \ or \Zboson \ boson in proton-proton collisions at $\sqrt{s}$ = 13 TeV}, Submitted to JHEP \\ (November 7, 2017), CMS-PAS-TOP-17-005 [\href{https://arxiv.org/abs/1711.02547}{\tt arXiv:1711.02547}].

\bibitem{ATLAS10}
ATLAS Collaboration, \emph{Measurement of the \ttbar\Zboson \ and \ttbar\Wboson \ production cross sections in multilepton final states using 3.2 \ifb \ of pp collisions at $\sqrt{s}$ = 13 TeV with the ATLAS detector}, \\ Eur. Phys. J. C {\bf 77} (2017) 40 [\href{https://arxiv.org/abs/1609.01599}{\tt arXiv:1609.01599v2}].

\bibitem{ATLAS14}
ATLAS Collaboration, \emph{Search for four-top-quark production in final states with one charged lepton and multiple jets using 3.2 \ifb \ of proton-proton collisions at $\sqrt{s}$ = 13 TeV with the ATLAS detector at the LHC} [\href{http://cdsweb.cern.ch/record/2144537}{\tt ATLAS-CONF-2016-020}].

\bibitem{CMS39}
CMS Collaboration, \emph{Search for standard model production of four top quarks in proton-proton collisions at $\sqrt{s}$ = 13 TeV}, Phys. Lett. B {\bf 772} (2017) 336 [\href{https://arxiv.org/abs/1702.06164}{\tt arXiv:1702.06164v2}].

\bibitem{CMS40}
CMS Collaboration, \emph{Search for standard model production of four top quarks with same-sign and multilepton final states in proton-proton collisions at $\sqrt{s}$ = 13 TeV}, Eur. Phys. J. C {\bf 78} (2018) 140 [\href{https://arxiv.org/abs/1710.10614}{\tt arXiv:1710.10614v2}].


\bibitem{ATLAS17}
ATLAS Collaboration, \emph{A measurement of the soft-drop jet mass in pp collisions at $\sqrt{s}$ = 13 TeV with the ATLAS detector}, Submitted to Phys. Rev. Lett. (November 24, 2017) [\href{https://arxiv.org/abs/1711.08341}{\tt arXiv:1711.08341v2}].

\bibitem{CMS32}
CMS Collaboration, \emph{Measurement of jet substructure observables in \ttbar \ events from pp collisions at \\ $\sqrt{s}$ = 13 TeV} (March 2018) [\href{http://cms-results.web.cern.ch/cms-results/public-results/preliminary-results/TOP-17-013/index.html}{\tt CMS-PAS-TOP-17-013}].

\bibitem{CMS33}
CMS Collaboration, \emph{Study of the underlying event in top quark pair production at 13 TeV} \\ (March 2018) [\href{http://cms-results.web.cern.ch/cms-results/public-results/preliminary-results/TOP-17-015/index.html}{\tt CMS-PAS-TOP-17-015}].

\bibitem{ATLAS18}
ATLAS Collaboration, \emph{Search for diboson resonances with boson-tagged jets in pp collisions at \\ $\sqrt{s}$ = 13 TeV with the ATLAS detector}, Phys. Lett. B {\bf 777} (2017) 91 [\href{https://arxiv.org/abs/1708.04445}{\tt arXiv:1708.04445v3}].

\bibitem{ATLAS19}
ATLAS Collaboration, \emph{Search for heavy resonances decaying to a \Wboson \ or \Zboson \ boson and a Higgs boson in the $\qqbar^{(\prime)}\bbbar$ final state in pp collisions at $\sqrt{s}$ = 13 TeV with the ATLAS detector}, \\ Phys. Lett. B {\bf 774} (2017) 494 [\href{https://arxiv.org/abs/1707.06958}{\tt arXiv:1707.06958}].

\bibitem{ATLAS20}
ATLAS Collaboration, \emph{Search for heavy resonances decaying into a \Wboson \ or \Zboson \ boson and a Higgs boson in final states with leptons and b-jets in 36 \ifb of $\sqrt{s}$ = 13 TeV pp collisions with the ATLAS detector}, JHEP {\bf 03} (2018) 174 [\href{https://arxiv.org/abs/1712.06518}{\tt arXiv:1712.06518v2}].















\end{thebibliography}
\end{document}

\bibliographystyle{JHEP} 